\documentclass[a4paper,fleqn]{cas-dc}

\usepackage[numbers]{natbib}
\usepackage{amsmath,amsfonts}
\usepackage{algorithmic}
\usepackage{algorithm}
\usepackage{array}
\usepackage{textcomp}
\usepackage{stfloats}
\usepackage{url}
\usepackage{verbatim}
\usepackage{graphicx}
\usepackage{comment}
\usepackage{caption}
\usepackage{adjustbox}
\usepackage{longtable}
\usepackage{subcaption}
\usepackage{orcidlink}
\usepackage{xurl}
\usepackage{hyperref}

\def\tsc#1{\csdef{#1}{\textsc{\lowercase{#1}}\xspace}}
\tsc{WGM}
\tsc{QE}
\tsc{EP}
\tsc{PMS}
\tsc{BEC}
\tsc{DE}

\begin{document}
\let\WriteBookmarks\relax
\def\floatpagepagefraction{1}
\def\textpagefraction{.001}

\shorttitle{Chlorophyll-a Mapping and Prediction in the Mar Menor Lagoon}
\shortauthors{A. Martínez-Ibarra et~al.}

\title [mode = title]{Chlorophyll-a Mapping and Prediction in the Mar Menor Lagoon Using C2RCC-Processed Sentinel 2 Imagery}

\nonumnote{This work is supported by project NEREIDAS TSI-100120-2024-13 funded by EU NextGenerationEU (PRTR). The study is also supported by grant RYC2023-043553-I, funded by MICIU/AEI/10.13039/501100011033 and ESF++, by the HORIZON-MSCA-2021-SE-01-01 project Cloudstars (g.a. 101086248), and partially funded by the Ministerio de Ciencia, Innovación y Universidades y la Agencia Estatal de Investigación (project AIA2025-163560-C41 (BLUEAI-UMU) funded by MICIU /AEI/10.13039/501100011033).}


\author[1]{Antonio Martínez-Ibarra}[orcid=0000-0001-6116-6811]
\cormark[1]
\ead{antonio.martinezi@um.es}

\author[1]{Aurora González-Vidal}[orcid=0000-0002-4398-0243]

\author[1]{Adrián Cánovas-Rodríguez}[orcid=0009-0002-7986-5615]

\author[1]{Antonio F. Skarmeta}[orcid=0000-0002-5525-1259]

\affiliation[1]{organization={Department of Information and Communication Engineering, University of Murcia},
                city={Murcia},
                postcode={30100},
                country={Spain}}

\cortext[cor1]{Corresponding author}


\begin{abstract}
The Mar Menor, Europe’s largest hypersaline coastal lagoon, located in southeastern Spain, has undergone severe eutrophication crises, with devastating impacts on biodiversity and water quality. Monitoring chlorophyll-a, a proxy for phytoplankton biomass, is essential to anticipate harmful algal blooms and guide mitigation strategies. Traditional \textit{in situ} measurements, while precise, are spatially and temporally limited. Satellite-based approaches provide a more comprehensive view, enabling scalable and long-term monitoring. This study aims to overcome limitations of chlorophyll monitoring, often restricted to surface estimates or limited temporal coverage, by developing a reliable methodology to predict and map chlorophyll-a concentrations across the water column of the Mar Menor. This work integrates Sentinel 2 imagery with buoy-based ground truth to create models capable of high-resolution, depth-specific monitoring, enhancing early-warning capabilities for eutrophication. Nearly a decade of Sentinel 2 images were atmospherically corrected using C2RCC processors. Buoy data were aggregated by depth (0–1 m, 1–2 m, 2–3 m, 3–4 m). Multiple machine learning algorithms, including CatBoost, XGBoost, Support Vector Machines, and Multilayer Perceptron Networks, were trained and validated using a cross-validation scheme with multi-objective optimization functions. Band-combination experiments and spatial aggregation strategies were tested to optimize prediction. The results show depth-dependent performance. The Root Mean Squared Logarithmic Error (RMSLE) obtained ranges from 0.34 at the surface to 0.39 at 3–4 m, while the $R^2$ value was 0.76 at the surface, 0.76 at 1–2 m, 0.70 at 2–3 m, and 0.60 at 3–4 m. Generated maps successfully reproduced known eutrophication events (e.g., 2016 crisis, 2025 surge), confirming robustness. The study delivers an end-to-end, validated methodology chlorophyll mapping. Its integration of multispectral band combinations, buoy calibration, and modeling offers a transferable framework for other turbid coastal systems.
\end{abstract}


\maketitle

\section{Introduction}

The Mar Menor, located in southeastern Spain's Region of Murcia, is Europe's largest hypersaline coastal lagoon. It covers an area of approximately 135 km² and has shallow waters with a maximum depth of seven meters. It has historically been characterized by clear waters, high salinity, and an oligotrophic nature. The lagoon is recognized as a unique and ecologically valuable system and is protected under multiple international designations \cite{SPAMIsRegionalActivity, EUNISSiteFactsheet}. However, since the mid-20th century, the lagoon has experienced increasing environmental degradation due to nutrient inputs from intensive agriculture, urban expansion, and former mining activity in the watershed. This chronic pollution has led to the progressive accumulation of nutrients, resulting in the first major eutrophication crisis in 2016 \cite{InformeIntegralSobre2017}. The resulting phytoplankton bloom, known as the ``green soup",  caused widespread turbidity and the collapse of submerged vegetation, marking a turning point in the system's ecological trajectory. Subsequent episodes of hypoxia and mass mortality of aquatic fauna, particularly in 2019 and 2021, further confirmed the lagoon’s loss of resilience. Even though there was some recovery in chlorophyll levels from 2022 to 2024, a new surge was detected in the summer of 2025 \cite{ColapsoMarMenor}. Chlorophyll, particularly chlorophyll-a (Chl-a), is essential for phytoplankton photosynthesis and is commonly used as an indirect indicator of its biomass, as it modifies the optical properties of water and gives it green tones. Concentrations of Chl-a reached 4-5 mg/m³ in the 2025 event, along with increased water turbidity and raising renewed concerns about the risk of anoxia and eutrophication. The recurrence and unpredictability of these eutrophication episodes underscore the need for improved monitoring strategies, as traditional approaches often lack the spatial and temporal resolution required to capture rapid ecological changes \cite{lopez-andreuMonitoringSystemMar2022}. 

In recent decades, the IoT paradigm has undergone significant expansion, especially in the field of environmental monitoring \cite{aryaGreenerTomorrowIoTEnabled2023, omkarbhagwankhilariIoTBasedEnvironment2024}. The proliferation of connected sensors, capable of collecting data in real time and transmitting it remotely, is transforming the way natural ecosystems are observed and managed. This trend responds to the need for denser, automated and continuous observation systems that allow early detection of environmental changes with high spatiotemporal resolution.

Particularly in vulnerable contexts, the deployment of IoT sensor networks has become a key tool to ensure adaptive, data-driven management. After the 2016 crisis, a network of multiparametric buoys was deployed in the Mar Menor, allowing the measurement of key water quality parameters, such as temperature, salinity, dissolved oxygen, and Chl-a. However, this approach faces operational challenges, such as sensor maintenance and biofouling. In addition, other sensors were deployed in nearby watercourses to measure streamflow and precipitation \cite{VisorSAIH}. Based on those, recent studies have also explored AI-based approaches to predict hydrological dynamics in the Mar Menor basin, with a focus on short-term streamflow forecasting in the Albujón watercourse \cite{cisterna-garciaArtificialIntelligenceStreamflow2025}.

Complementing \textit{in situ} instrumentation, remote sensing has become a valuable tool for the spatial and temporal monitoring of water quality in the Mar Menor, particularly Chl-a. Parameters such as Chl-a concentration, (an indicator of phytoplankton biomass, and turbidity have optical properties that allow them to be studied using satellite images. Indeed, algal pigments alter the reflectance of water at certain wavelengths \cite{parsonsManualChemicalBiological1984}, so that multispectral sensors on board satellites can be used to estimate surface Chl-a concentration from reflected radiation. This capability offers notable advantages: unlike buoys, satellite images cover the entire area of the lagoon, allowing the spatial distribution of algal blooms to be mapped and the most affected areas to be detected. In addition, modern satellites have orbits that span a few days, which facilitates more frequent monitoring. 
Therefore, the use of satellite imagery provides a synergistic monitoring system alongside direct measurements: while buoys provide data at fixed points, remote sensing provides a synchronous view of the entire lagoon. This multi-parametric and multi-scale integration is especially useful in a fragile ecosystem such as the Mar Menor, as it facilitates the early detection of algal blooms and the monitoring of their evolution.

The objective of this study is to create a reliable model of Mar Menor's Chl-a concentration using satellite imagery to improve spatial and temporal resolution for consistent monitoring. To that end, different sources of data are used as input: \textit{in situ} data from buoys and satellite images gathered from the European Space Agency \cite{portaldelaagenciaespacialeuropeaESASpain}. Once a set of reliable, tested models is developed and actively working, Mar Menor monitoring will be more comprehensive, and the output chlorophyll maps will help achieve the goal of making data more accessible to the public.

To summarize, the main contributions of this study are:
\begin{itemize}
    \item An in-depth exploration of multispectral band combinations as predictors evaluated with a wide array of machine learning (ML) models.
    \item An end-to-end methodology for predicting Chl-a in turbid waters environments, leveraging the Case 2 Regional CoastColor (C2RCC) processors for atmospheric correction of Sentinel 2 imagery and using buoy data as ground truth to train the models. The methodology has been validated both quantitatively, through performance metrics, and qualitatively, by reproducing known states of the lagoon.
    \item The capability to generate Chl-a maps not only at the surface but also throughout the water column, providing a framework to study the lagoon’s evolution since the beginning of Sentinel 2 records.
    \item A pipeline for automatic chlorophyll map generation that, starting from a selected date, processes Sentinel 2 SAFE files in SNAP, extracts reflectances for all pixels in the Mar Menor, incorporates additional features, such as band combinations, and performs model inference to generate chlorophyll maps in a fully automated manner. Available in the public Github repository of this project\footnote{\url{https://github.com/Antonio-MI/mar-menor-chl}}.
\end{itemize}

The structure of the paper is as follows: Section \ref{sec:background} describes the background knowledge needed to contextualize and understand this study and discusses other works that use remote sensing in similar scenarios. Then, in Section \ref{sec:materials} the study area is described, along with the satellite imagery and buoy data. Section \ref{sec:methodology} details the procedure carried out to gather, process and model the data. Section \ref{sec:exp} describes the experiments and results, and finally Section \ref{sec:conc} highlights the outcomes of this study and proposes future lines of work.

\section{Background and Related Work}\label{sec:background}

This section includes details about the Mar Menor situation, the Sentinel 2 constellation and discusses related works in this field.

\subsection{Background}

\subsubsection{Mar Menor}

The Mar Menor, located in the Region of Murcia, in southeastern Spain, is the largest hypersaline coastal lagoon in Europe, and until a few years ago, it was characterized by its crystalline waters, high salinity and oligotrophic nature (very low nutrients disolved), which made it a unique ecosystem of great environmental value. Actually, the ecological importance of the Mar Menor has been recognized with multiple protection figures (regional, state, European and international), including its designation as a Specially Protected Area of Mediterranean Importance (Barcelona Convention) \cite{SPAMIsRegionalActivity}, as well as a Site of Community Importance and a Special Protection Area for Birds within the Natura 2000 Network \cite{EUNISSiteFactsheet}.

Despite its natural value, the lagoon has suffered a gradual deterioration over the last decades. Since the mid-20th century, there has been intense socioeconomic development in its watershed with the expansion of irrigated agriculture, intensive livestock farming, urban-tourist growth and former mining operations nearby. These activities have exerted sustained polluting pressures for decades, discharging nutrients such as nitrates and phosphates, among others, into the lagoon through seasonal watercourses and the subway aquifer, leading to a progressive eutrophication of the ecosystem. In fact, over the last 40 years the Mar Menor has suffered chronic pollution from agricultural runoff, which has led to large-scale algal blooms and marked ecological degradation. It should be noted that, despite signs of degradation, until well into the 21st century some environmental indicators still showed low nutrient and chlorophyll values, which led to the lagoon being perceived as relatively oligotrophic. However, nutrients continued to accumulate in the system, making it increasingly vulnerable to ecological collapse \cite{ColapsoMarMenor}.

The turning point came in 2016 \cite{ProblematicaActual}, the year in which the Mar Menor suffered its first major eutrophic crisis. During the spring-summer of that year, the phenomenon known as ``green soup" took place: an explosive growth of phytoplankton (microalgae) fed by the excess of nutrients in the water column. The algal proliferation was so dense that it completely clouded the waters and prevented the passage of light beyond ~3 m depth, causing the mortality of most of the underwater meadows of phanerogams and macroalgae at the bottom of the lagoon. This event marked an unprecedented ecological collapse in the Mar Menor, showing that the nutrient load had exceeded the system's capacity for self-regulation. During this crisis, Chl-a concentrations peaked over 30 mg/m$^3$. To put this value into perspective, in 2002 a value higher than 7 mg/m$^3$ was considered a maximum \cite{perez-ruzafaSpatialTemporalVariations2005}. 
According to \cite{sheelaEnvironmentalStatusTropical2011}, a water body is considered oligotrophic when Chl-a concentrations range between 0 and 2.6 mg/m$^3$, mesotrophic between 2.6 and 7.3 mg/m$^3$, and eutrophic above 7.3 mg/m$^3$. Accordingly, in 2002 the lagoon was close to the eutrophic threshold, whereas in 2016 it exceeded it substantially.

In subsequent years, new critical episodes associated with eutrophication occurred. In autumn 2019, following torrential rains produced by a DANA (an isolated high-altitude atmospheric depression that can generate extreme precipitation events in the Mediterranean region), an episode of anoxia was triggered in the waters causing massive mortality rates on fish and crustaceans. This was due to the high concentration of phytoplankton in water surface, keeping light from reaching the bottom and thus preventing photosynthesis to take place, leading to reduced amounts of dissolved oxygen whcih led to caused a massive mortality of fish and crustaceans.
More recently, in August 2021, another acute hypoxia event was recorded in several areas of the lagoon, again linked to an exacerbated phytoplankton bloom. This bloom of 2021, originating around the mouth of the Albujón seasonal watercourse, caused again the suffocation of the aquatic biota and the consequent mortality of thousands of organisms. During 2022, 2023 and 2024 chlorophyll values returned to typical values around 0.5-1.0 mg/m³, but despite various management and mitigation efforts since then, the lagoon remains in a fragile state. Proof of this is the situation currently recorded \cite{pressMarMenorEntra2025}, in the summer of 2025, when a worrying rebound in Chl-a has been detected. Chl-a levels in the water have returned to peaks of 4-5 mg/m³. This increase in Chl-a, observed from July 2025, has been accompanied by an increase in turbidity, once again placing the Mar Menor on alert for the risk of another episode of anoxia if the trend persists.

The severe environmental crises experienced, especially since 2016, have highlighted the need to monitor the conditions of the Mar Menor more closely to anticipate critical episodes \cite{erena2017analisis}. In that year, a network of in-situ sensors was implemented, consisting of multiparametric oceanographic buoys distributed throughout the lagoon. Each buoy is equipped with sensors that measure conductivity, temperature, dissolved oxygen, and Chl-a at different depths, but with limited time resolution. Additionally, several campaigns to take \textit{in situ} measurements have been carried out, obtaining also values for the aforementioned parameters. This intensive monitoring of the Mar Menor represents a significant advance towards its protection, although it is not enough if the causes of eutrophication in its basin are not addressed. The dependence on buoys also poses a problem: they require frequent on-site maintenance to prevent the sensors from being covered by algae, becoming filled with salt, etc.

\subsubsection{Sentinel 2}

From all the available satellite imagery resources, Sentinel 2 was chosen for this study due to its high resolution and sample frequency, but the methodology proposed here can be further extended to  satellites like Landsat, Sentinel-3 or Modis, among others.

The Sentinel 2 mission is part of the Copernicus program of the European Union, and consists of an optical satellite constellation (Sentinel 2A since 2015, 2B since 2017, and, starting in 2024, 2C) designed to provide high-resolution multispectral images with global coverage every five days at the equator (every two or three days in mid-latitudes), covering terrestrial, coastal, and Mediterranean Sea areas between 56° south and 84° north latitude.

Each satellite carries an MSI instrument (Multispectral Instrument) that captures thirteen spectral bands, from visible to near-infrared and shortwave infrared, detailed in Table \ref{table:Sentinel2Bands}. The mission provides images with spatial resolutions of 10 m (4 bands), 20 m (6 bands), and 60 m (3 bands), with a field of view of approximately 290 km \cite{europeanspaceagencyS2Mission}.

Among its products, two are the most used: Level 1C (L1C) and Level 2A (L2A).

The Sentinel 2 L1C product provides orthorectified images in Top of Atmosphere (TOA) reflectance, i.e. solar radiation reflected back to the satellite without removing atmospheric effects. These products are georeferenced in the UTM/WGS84 system and are organized in regular 100 km $\times$ 100 km mosaics, maintaining the native spatial resolution of each band (10, 20 or 60 meters). They include basic geometric and radiometric corrections, as well as various masks that indicate saturated pixels, areas without data or anomalies in the image. Since they do not incorporate atmospheric correction, the L1C products are especially useful for users who wish to apply proprietary algorithms or perform custom processing from raw data, although these products are not directly comparable between dates without additional processing \cite{europeanspaceagencyS2Products}.

The L2A product represents a more advanced and directly usable version, as it includes Bottom of Atmosphere (BOA) type reflectance, i.e. the signal reflected from the land or water surface after removing the effects of the atmosphere using the Sen2Cor correction processor  \cite{sentinelonlineCollection0Level2A}. It is also delivered in the same map geometry, UTM, and with the bands in their native resolution. These products also incorporate a pixel classification layer known as Scene Classification Map (SCL), which identifies areas of cloud, cloud shadow, vegetation, bare ground, water or other types of cover, facilitating the filtering of invalid or contaminated data. L2A products are commonly used in environmental applications such as monitoring water bodies, vegetation or land use dynamics.


\begin{table}[!t]
    \caption{Sentinel 2 bands}
    \label{table:Sentinel2Bands}
    \centering
    \resizebox{0.49\textwidth}{!}{%
    \begin{tabular}{|p{0.9cm}|p{1.3cm}|p{1.6cm}|p{3.8cm}|}
        \hline
        \textbf{Band} & \textbf{Resolution} & \textbf{Central Wavelength} & \textbf{Description} \\ \hline
        B1    & 60 m  & 443 nm   & Ultra Blue (Coastal and Aerosol) \\ \hline
        B2    & 10 m  & 490 nm   & Blue \\ \hline
        B3    & 10 m  & 560 nm   & Green \\ \hline
        B4    & 10 m  & 665 nm   & Red \\ \hline
        B5    & 20 m  & 705 nm   & Visible and Near Infrared (VNIR) \\ \hline
        B6    & 20 m  & 740 nm   & Visible and Near Infrared (VNIR) \\ \hline
        B7    & 20 m  & 783 nm   & Visible and Near Infrared (VNIR) \\ \hline
        B8    & 10 m  & 842 nm   & Visible and Near Infrared (VNIR) \\ \hline
        B8a   & 20 m  & 865 nm   & Visible and Near Infrared (VNIR) \\ \hline
        B9    & 60 m  & 940 nm   & Short Wave Infrared (SWIR) \\ \hline
        B10   & 60 m  & 1375 nm  & Short Wave Infrared (SWIR) \\ \hline
        B11   & 20 m  & 1610 nm  & Short Wave Infrared (SWIR) \\ \hline
        B12   & 20 m  & 2190 nm  & Short Wave Infrared (SWIR) \\ \hline
    \end{tabular}}
\end{table}

Satellite images need atmospheric corrections in order to obtain reliable reflectances across all the wavelengths measured. The atmospheric correction performed by the L2A product is not enough to obtain an useful product, since water reflectances make the problem highly complicated. One of the main challenges in the study of inland and coastal water bodies by remote sensing is accurately correcting the reflectance measured by satellite sensors for atmospheric effects. This challenge is exacerbated in waters classified as Case 2, which are rich in suspended particles, dissolved organic matter, and phytoplankton. The optical properties of these waters are highly complex and variable. The C2RCC (Case 2 Regional CoastColor) processor \cite{doerfferMERISCase22007, c2rcccommunityprojectDocumentationC2RCC} offers a solution consisting of neural networks that are trained using an extensive database of radiative transfer simulations. These networks perform atmospheric correction and estimate the Inherent Optical Properties (IOPs) of water from TOA reflectances. This is needed because at satellite altitude most of sensor-measured signal comes from the surface and the atmosphere \cite{wangAtmosphericCorrectionOcean2014}.

The C2RCC processor excels in its ability to operate in optically complex environments thanks to a bio-optical model that better captures the diversity of optical conditions in inland and coastal waters. Neural networks are trained to invert TOA reflectances and recover water-leaving reflectance. One of the great advantages of C2RCC over other more empirical or semi-analytical approaches is its ability to process images from multiple sensors (Sentinel 2 MSI, Sentinel-3 OLCI, MODIS, MERIS, etc.) without retraining, making it an operational, scalable and flexible tool for environmental monitoring \cite{brockmannEvolutionC2RCCNeural2016}. 

To extend the applicability of the C2RCC processor to a wider range of optical conditions in inland and coastal waters, variants of the processor based on neural networks trained with different ranges of IOPs have been developed \cite{soriano-gonzalezCombinationC2RCCProcessors2022}. The original version, C2RCC-Net, is optimized for medium-turbidity coastal waters. The C2X-Net version considerably expands these training ranges by incorporating extreme cases of suspended matter concentration and absorption by pigments and dissolved organic matter. This was accomplished by supplementing the original database with additional simulations from the CoastColour project, including highly eutrophic water conditions and limiting cases. C2X-Net has demonstrated an enhanced capacity to accurately recover water reflectance in eutrophic lakes, turbid estuaries, and intricate inland lagoons, environments in which more conservative algorithms often falter or produce systematic biases.
An intermediate version, C2X-Complex (C2XC), was subsequently introduced to balance the sensitivity of C2RCC with the robustness offered by C2X in extreme conditions \cite{soriano-gonzalezCombinationC2RCCProcessors2022}. C2XC allows for positive results in moderately turbid waters and more productive bodies without the overfitting or loss of spectral accuracy issues that can occur with C2X in clearer scenarios.

Therefore, C2RCC and its variants are a particularly useful tool for monitoring vulnerable water bodies, such as the Mar Menor, where multitemporal remote sensing of Chl-a can provide information on critical eutrophication events, their spatial magnitude and temporal evolution.

The processor is available as an integrated module in the Sentinel Application Platform (SNAP) software developed by ESA \cite{SNAPEarthOnline}, allowing its direct application on Sentinel 2 images with an accessible and user-configurable interface.

\subsection{Related Work}

Since the 2016 eutrophication crisis more options for monitoring water quality have been explored. Once the Sentinel 2 constellation, consisting of 2A and 2B, became operational in 2017, remote sensing techniques became much more viable. Prior to that, \cite{erenaAnalisisMultisensorVariabilidad2017} studied the use of Landsat and Spot satellites to understand the variability and evolution of Chl-a in the Mar Menor. This study examines the evolution of Chl-a in the Mar Menor using \textit{in situ} and remote sensing data from multispectral sensors on the Landsat 8 Operational Land Imager (OLI) and SPOT 7 satellites. Normalized indexes were used between bands of the OLI sensor in the visible and near-infrared range of the type: $$\frac{R(\lambda_1) - R(\lambda_2)}{R(\lambda_1) + R(\lambda_2)}$$ where $R(\lambda_i)$ corresponds to the different band reflectances. During the study period from May 2015 to May 2017, the best results were achieved using the index normalized with the green and blue bands, reaching an R² of 0.88. Further work from the same authors \cite{erenaMonitoringCoastalLagoon2019} focused on obtaining maps using the empirical algorithm previously proposed, and they also studied the relationship between Chl-a and turbidity. The study revealed the intricate variability between these parameters, as they were not consistently correlated due to factors such as atmospheric conditions, water column movements, and phytoplankton distribution.

In 2020, \cite{jimeno-saezUsingMachineLearningAlgorithms2020} used ML models to predict the concentration of Chl-a in the Mar Menor lagoon based on physical and chemical water data. Two approaches were evaluated: multilayer neural networks and support vector regression (SVR). Parameters such as water temperature, pH, suspended solids, and turbidity were used as predictor variables. However, this study does not use remote sensing or spectral bands; it is based exclusively on \textit{in situ} measured parameters, so there are no spectral formulas or satellite image use. Yet, the study emphasizes that ML models are especially useful for monitoring this type of environment.

The literature regarding remote sensing with Sentinel 2 for Mar Menor starts to proliferate in 2021. \cite{gomezNewApproachMonitor2021} proposed a methodology using ML and deep learning (DL) models to estimate surface (less than 0.5 m) Chl-a. The inputs included band combinations such as the normalized difference chlorophyll index (NDCI), the normalized difference turbidity index (NDTI), and $B3/B8$ ratios, among others. Images are resampled up to 60 meters, and the most important features in the optimal model are the $B3/B8$ ratio, NDTI, and bands 9 and 2. Additionally, feature selection was explored in three scenarios: all variables, removal of features with correlations greater than 0.75, and Principal Component Analysis (PCA). Overall, maintaining all variables yielded the best results.

\cite{caballeroUseSentinel2Landsat82022} combined the two  previously mentioned satellites: Sentinel 2 and Landsat 8. Their study focused on a brief period of time in 2021 and they implemented atmospheric corrections with ACOLITE, a toolbox designed to correct Level 1 products over marine, inland and coastal waters. Then, they estimated Chl-a using the bio-optical algorithm OC3 \cite{MODISChlorophyllNOAA} based on water inherent properties and applied the model to the whole lagoon. In the case of \cite{lopez-andreuMonitoringSystemMar2022}, rather than applying the model to each pixel to map the entire water surface, the best performing model was applied only to points where \textit{in situ} measurements were taken. Then, an interpolation method such as distance weighting or kriging was used to map the rest of the water surface.

In \cite{yeleyMarMenorLagoon2022}, SNAP was used to correct Sentinel 2 images with the set C2X of C2RCC. With the atmospherically corrected images, reflectances are grouped within a 3 x 3 pixel window. Then, Chl-a is estimated using empirical models in which different band combinations are tested to fit the line $y = ax + b$. The band combination that yielded the best results is:
$$\frac{B3 + B5}{B3 + B4}$$
The fitted line is $y = 124.94x - 115.35$, with an R squared value of 0.82. To approach our study, we first replicated this one with the same dates. With our data, we achieved an R-squared value close to 0.8, which is quite close to the results of the article. Further research shows that this value is mainly due to one image: 2019-09-18. This image has 40.53\% cloud cover, meaning some of the data gathering stations are covered by clouds. Furthermore, this date is the only one with chlorophyll values greater than 10 mg/m³ in the dataset. Excluding that image from the same procedure leads to an R-squared value close to 0. Nevertheless, the proposed methodology provided a foundation for this work.

While previous studies focused only on surface chlorophyll (less than one meter), \cite{gimenezEnhancingShallowWater2024} explored prediction across the entire water column. They used several ML and DL models along with Sentinel-3 imagery to predict Chl-a concentrations at various depths for each \textit{in situ} measurement point. They used 21 OLCI spectral bands as input and feature selection based on statistical significance. The best results are achieved on the surface, and as the depth increases, metrics deteriorate.

The most recent study in the Mar Menor region \cite{gomez-jakobsenMonitoringChlorophyllConcentration2025} focused on using MODIS, VIIRS, and Sentinel-3 sensors to monitor Chl-a. The study developed an empirical algorithm named BELA based on the red-to-green ratio, which is expressed by the following formula: $Chl = 0.353 RG^{3} + 2.132RG^{2} + 3.905RG + 2.110$ where $RG = \log_{10}(R_{670}/R_{550})$. However, when we applied this relationship to Sentinel 2 reflectances and our data, the results were not satisfactory. This may be due to the significant differences between the sensors in this article and the MSI, as well as the atmospheric corrections we used.

Remote sensing has been also widely used in other areas with turbid and productive water bodies. \cite{mishraNormalizedDifferenceChlorophyll2012} introduced the NDCI to estimate Chl-a in Mobile Bay and the Mississippi Delta with MERIS imagery adjusting the model with simulated data to cover a wide range of chlorophyll values. Building on this, \cite{pahlevanSeamlessRetrievalsChlorophylla2020} also employed NDCI in combination with probabilistic-based networks, arguing that chlorophyll is only one of several components influencing reflectance. \cite{wattelezStatisticalAlgorithmEstimating2016} uses MODIS and proposes the idea of using a polynomial with logarithms for low values of Chl-a and a SVM for Chl-a greater than 3 mg/m$^3$ while \cite{linsAssessmentChlorophyllaRemote2017} explores the use several satellites in a tropical estuarine lagoon in Brasil. Finally, focusing on the Ebro Delta in Spain, \cite{fernandez-tejedorAccurateEstimationChlorophylla2022} applied the C2RCC and C2X processors for atmospheric correction of Sentinel 2 imagery and tested several band combinations using linear, simple polynomial, and logarithmic models for chlorophyll estimation.

Table \ref{tab:related} summarizes all the articles discussed in a compressed format, so the reader can easily compared the methods used, band combinations, chlorophyll ranges, etc.

\begin{table*}[t!]
\caption{Related Work summary table. Units for RMSE and MAE are mg/m$^3$ }
\label{tab:related}
\centering
\resizebox{\textwidth}{!}{%
\begin{tabular}{|l|l|l|l|l|l|l|l|l|l|l|}
\hline
\textbf{Paper}                                              & \textbf{Area}                                                              & \textbf{Satellite}                                                                    & \textbf{\begin{tabular}[c]{@{}l@{}}Resolution \\ (m)\end{tabular}} & \textbf{Band combinations}                                                        & \textbf{Date Range}                                            & \textbf{Models}                                                                     & \textbf{\begin{tabular}[c]{@{}l@{}}Chl Range \\ (mg/m³)\end{tabular}} & \textbf{\begin{tabular}[c]{@{}l@{}}Predicted\\ Depth\end{tabular}}     & \textbf{Metrics}                                                                         & \textbf{\begin{tabular}[c]{@{}l@{}}Train/\\ Test\end{tabular}} \\ \hline
\cite{erenaMonitoringCoastalLagoon2019}                     & Mar Menor                                                                  & Landsat 8                                                                             & 30                                                                 & \begin{tabular}[c]{@{}l@{}}Normalized difference \\ (Green, Blue)\end{tabular}    & \begin{tabular}[c]{@{}l@{}}May 2015 - \\ May 2017\end{tabular} & \begin{tabular}[c]{@{}l@{}}Manual adjusment \\ (exp function)\end{tabular}          & 5 - 25                                                                & 0 - 1 m                                                                & R² = 0.88                                                                                & No                                                             \\ \hline
\cite{gomezNewApproachMonitor2021}                          & Mar Menor                                                                  & Sentinel 2                                                                            & 60                                                                 & \begin{tabular}[c]{@{}l@{}}NDCI, NDTI, B3/B8,\\  B1 to B12\end{tabular}           & \begin{tabular}[c]{@{}l@{}}May 2017 - \\ Jan 2019\end{tabular} & RF, SVR, ANN, MLP                                                                   & 0 - 13                                                                & 0 - 1 m                                                                & \begin{tabular}[c]{@{}l@{}}R² = 0.92 (RF)\\ RMSE = 0.82 (RF)\end{tabular}                & Yes                                                            \\ \hline
\cite{caballeroUseSentinel2Landsat82022}                    & Mar Menor                                                                  & \begin{tabular}[c]{@{}l@{}}Sentinel 2 + \\ Landsat 8\end{tabular}                     & 10                                                                 & B1 to B12                                                                         & \begin{tabular}[c]{@{}l@{}}Mar 2021 - \\ Nov 2021\end{tabular} & OC3 (bio-optical)                                                                   & 0.5 - 5                                                               & 0 - 1 m                                                                & \begin{tabular}[c]{@{}l@{}}R² = 0.90\\ MAE = 0.43\end{tabular}                           & No                                                             \\ \hline
\cite{lopez-andreuMonitoringSystemMar2022}                  & Mar Menor                                                                  & \begin{tabular}[c]{@{}l@{}}Sentinel 2 + \\ Landsat 8\end{tabular}                     & 10                                                                 & -                                                                                 & \begin{tabular}[c]{@{}l@{}}Jan 2021 - \\ Jul 2022\end{tabular} & RF, LBM, XGB                                                                        & -                                                                     & \begin{tabular}[c]{@{}l@{}}Average\\ of 0 to 5 m\end{tabular}          & -                                                                                        & No                                                             \\ \hline
\cite{yeleyMarMenorLagoon2022}                              & Mar Menor                                                                  & Sentinel 2                                                                            & 10                                                                 & \begin{tabular}[c]{@{}l@{}}(Green + NIR1)/\\ (Green + Red)\end{tabular}           & \begin{tabular}[c]{@{}l@{}}Oct 2016 - \\ Oct 2019\end{tabular} & LR                                                                                  & 0 - 28                                                                & 0 - 1 m                                                                & \begin{tabular}[c]{@{}l@{}}R² = 0.82\\ RMSE = 2.6\end{tabular}                           & No                                                             \\ \hline
\cite{gimenezEnhancingShallowWater2024}                     & Mar Menor                                                                  & Sentinel 3                                                                            & 300                                                                & 21 OLCI Bands                                                                     & \begin{tabular}[c]{@{}l@{}}Aug 2016 - \\ Feb 2022\end{tabular} & \begin{tabular}[c]{@{}l@{}}LR, DT, RF, KNN,\\  MLP, CNN\end{tabular}                & 0 - 28                                                                & \begin{tabular}[c]{@{}l@{}}0 - 1, 1 - 2,\\ 2 - 3, 3 - 4 m\end{tabular} & \begin{tabular}[c]{@{}l@{}}R² = 0.89 - 0.64 (CNN)\\ MAE = 0.84 - 1.25 (CNN)\end{tabular} & Yes                                                            \\ \hline
\cite{gomez-jakobsenMonitoringChlorophyllConcentration2025} & Mar Menor                                                                  & \begin{tabular}[c]{@{}l@{}}MODIS, \\ Sentinel 3, \\ VIIRS\end{tabular}                & 300 - 1000                                                         & Red/Green                                                                         & \begin{tabular}[c]{@{}l@{}}Jun 2016 - \\ Jun 2023\end{tabular} & \begin{tabular}[c]{@{}l@{}}BELA (third order \\ polynomial)\end{tabular}            & 2 - 30                                                                & 4 m                                                                    & R² = 0.78                                                                                & No                                                             \\ \hline
\cite{mishraNormalizedDifferenceChlorophyll2012}            & \begin{tabular}[c]{@{}l@{}}Mobile Bay and\\ Mississippi Delta\end{tabular} & MERIS                                                                                 & 300                                                                & NDCI                                                                              & -                                                              & \begin{tabular}[c]{@{}l@{}}Second order \\ polynomial\end{tabular}                  & 0 - 60                                                                & Not specified                                                          & \begin{tabular}[c]{@{}l@{}}R² = 0.94\\ RMSE = 1.43\end{tabular}                          & Yes                                                            \\ \hline
\cite{wattelezStatisticalAlgorithmEstimating2016}           & \begin{tabular}[c]{@{}l@{}}New Caledonian \\ Lagoon\end{tabular}           & MODIS                                                                                 & 500                                                                & \begin{tabular}[c]{@{}l@{}}Blue/Green, \\ UltraBlue/Green\end{tabular}            & -                                                              & \begin{tabular}[c]{@{}l@{}}Polynomial with \\ logarithms, \\ SVM, OC3\end{tabular}  & 0 - 38                                                                & 2 m                                                                    & RMSE = 0.67 (SVM)                                                                        & Yes                                                            \\ \hline
\cite{linsAssessmentChlorophyllaRemote2017}                 & \begin{tabular}[c]{@{}l@{}}Mundaú-\\ Manguaba\end{tabular}                 & \begin{tabular}[c]{@{}l@{}}MODIS, \\ MERIS, \\ Sentinel 2, \\ Sentinel 3\end{tabular} & 10 - 500                                                           & Blue/Green, NIR/Red                                                               & -                                                              & \begin{tabular}[c]{@{}l@{}}LR with two, three \\ and four band ratios\end{tabular}  & 5 - 117                                                               & 0 - 1 m                                                                & \begin{tabular}[c]{@{}l@{}}R² = 0.78\\ RMSE = 10.44\end{tabular}                         & Yes                                                            \\ \hline
\cite{fernandez-tejedorAccurateEstimationChlorophylla2022}  & Ebro Delta                                                                 & Sentinel 2                                                                            & 10                                                                 & \begin{tabular}[c]{@{}l@{}}Ratios (B2 to B6), \\ NDCI,\\ Three bands\end{tabular} & -                                                              & \begin{tabular}[c]{@{}l@{}}LR, simple and \\ logartihmic\\ polynomials\end{tabular} & 0 - 9                                                                 & \begin{tabular}[c]{@{}l@{}}Average \\ of 4 to 15 m\end{tabular}        & \begin{tabular}[c]{@{}l@{}}R² = 0.88 (2 deg polynomial)\\ MAE = 0.598\end{tabular}       & Yes                                                            \\ \hline
\end{tabular}
}
\end{table*}

While numerous studies have demonstrated the potential of remote sensing for monitoring Chl-a in turbid and optically complex waters, certain aspects remain underexplored, particularly in regions such as the Mar Menor. Open-source methodologies that integrate long-term datasets, cover a broad range of Chl-a concentrations, explore predictions across the whole water column, and include systematic comparisons of multiple modeling strategies are still relatively scarce. Moreover, although neural network-based tools such as C2RCC and C2X are increasingly used, their integration into customized regional workflows and their comparison with other modeling approaches are areas that merit further development.

In this study, we address these gaps by developing and openly releasing a methodology tailored for the retrieval of Chl-a in optically complex coastal waters, using nearly a decade of Sentinel 2 data at 10 m resolution. This approach systematically evaluates a wide range of band combinations and processing strategies. It integrates neural-network-based atmospheric corrections (C2RCC, C2X, and C2X-Complex) with an advanced suite of machine learning regressors, including CatBoost, XGBoost, SVR, and other ML algorithms, to optimize chlorophyll-a (Chl-a) estimation across the water column. This comprehensive analysis is conducted within a robust multi-objective optimization and cross-validation framework. Additionally, we provide clear guidance and visual examples--in Section \ref{sec:exp} and in the Github repository\footnote{\url{https://github.com/Antonio-MI/mar-menor-chl}}--of how the best-performing models can be operationally applied to generate Chl-a maps, contributing to improved transparency, reproducibility, and practical utility in regional water quality monitoring. 

\section{Materials and methods}\label{sec:materials}

This section describes the area where the study was made, the satellite imagery used, and the buoy data gathered.

\subsection{Study area}

The Mar Menor is a coastal lagoon separated from the Mediterranean Sea by La Manga, with an approximate area of 135 km$^2$ and shallow waters, less than 7 meters deep. The region's climate is classified as a dry Mediterranean climate, with an average annual temperature ranging from 17°C to 18°C. The summers are hot, with temperatures over 40ºC, and the winters are mild, with extremes in temperature mitigated by the influence of the nearby sea. The region experiences low rainfall, typically ranging from 270 millimeters to 350 millimeters per year \cite{MarcoFisico}. The temperature of the water body is subject to variation according to the season. During the winter months, the temperature is approximately 15°C. In the spring, the temperature typically ranges from 16–18°C. During the summer months, the temperature reaches an average of 24-25°C, and there have been recorded temperatures of 31-32°C during recent episodes.

The salinity of the Mar Menor typically varies between 40 and 47 Practical Salinity Unit (PSU), which is considered high for a coastal lagoon and attributed to minimal rainfall and substantial evaporation, and classifies it as an hypersaline coastal lagoon.
In hydrodynamic terms, the Mar Menor displays marked homogeneity in its internal behavior, with minor variations in level between sensors except during episodes of intense wind, where seiche-type oscillations are generated in the lagoon. The wind also generates a water level difference of up to 10 centimeters and directs surface currents in its direction, thereby facilitating water mixing and renewal \cite{InformeIntegralSobre2017}.

\subsection{Methods}

This subsection outlines the ML models that were implemented. This selection explores a wide range of algorithms, covering classical statistical approaches and modern ML techniques. Some of these models were selected because previous studies have already demonstrated their suitability for predicting Chl-a and related water quality parameters in coastal and inland waters. Others were included to expand upon those findings and assess whether additional modeling paradigms could improve robustness and generalization. This includes distance-based methods, linear and regularized regression models, tree-based learners, and gradient boosting frameworks. Additionally, a simple multilayer perceptron is also included, and ensemble strategies are used to combine individual predictors. This variety of models enables a thorough comparison of different learning paradigms and demonstrates their respective strengths in Chl-a prediction.

\textbf{Linear Regression (LR)} \\
A supervised learning method in which the output is modeled as a linear combination of input features, capturing the approximately linear relationships between Chl-a concentration and spectral reflectance \cite{matus2018predictive}. The model assumes a constant rate of change in the output for each unit change in the inputs, and its parameters are typically estimated by minimizing the sum of squared errors \cite{LinearRegressionML}. Implemented with LinearRegression from \texttt{sklearn}.

\textbf{K-Nearest Neighbors (KNN)} \\
A non-parametric regression method where the prediction for a given instance is based on the average of the target values of its $k$ closest neighbors in the feature space. In this work the weights are assigned proportionally to the inverse distance from the query point, and distance is computed as euclidean distance. KNN could suit this problem because it can capture local nonlinearities between reflectance and Chl-a concentration without assuming a specific functional form. Implemented with KNeighborsRegressor from \texttt{sklearn} \cite{KNeighborsRegressor}.

\textbf{Support Vector Regressor (SVR)} \\
A regression algorithm derived from Support Vector Machines (SVM) that seeks to fit a function within an error margin (an epsilon-insensitive tube) while penalizing points that fall outside this margin. SVR effectively captures complex nonlinear relationships between environmental factors and Chl-a concentration, avoids overfitting through structural risk minimization, performs well with small sample sizes, and provides high prediction accuracy even when data are noisy \cite{xu2023prediction}. Implemented using the SVR module from \texttt{scikit-learn} \cite{SVR}.

\textbf{ElasticNet (ELN)} \\
A linear regression model that combines L1 (lasso) and L2 (ridge) regularization penalties. It is useful when there are many correlated features: L1 helps with feature selection (sparse coefficients), L2 stabilizes the coefficient values. In this context, ELN helps reduce redundancy between highly correlated spectral bands while maintaining model stability. Implemented with ElasticNet from \texttt{sklearn} \cite{ElasticNet}.

\textbf{Random Forest (RF)}
An ensemble learning method that builds many decision trees on bootstrapped subsets of the data and averages their predictions to reduce overfitting and improve generalization. RF can handle nonlinear relationships, tolerate correlated variables, provide high prediction accuracy, and objectively rank the importance of environmental factors. Implemented with RandomForestRegressor from \texttt{sklearn} \cite{RandomForestRegressor}.

\textbf{Light Gradient Boosting Machine (LBM)} \\
A gradient boosting framework optimized for efficiency and scalability. It builds decision trees sequentially, each one correcting the errors of the previous ones, using histogram-based methods and leaf-wise tree growth to speed up training \cite{LightGBM}. In the context of Chl-a prediction with satellite imagery, LightGBM is particularly appealing because it can handle large volumes of pixel-level data efficiently, exploit subtle nonlinear interactions among spectral bands, and scale to the many candidate band combinations tested in our workflow. Implemented with LGMBRegressor from Microsoft's LightGBM.

\textbf{eXtreme Gradient Boosting (XGB)} \\
A high-performance gradient boosting implementation that uses second-order derivatives for optimization, and includes regularization to avoid overfitting. Known for speed, accuracy and flexibility in hyperparameter tuning. For Chl-a prediction in optically complex waters, XGBoost is a strong candidate since it can model nonlinear relationships between reflectance and Chl-a, manage high-dimensional feature spaces (e.g. many band combinations), and incorporate regularization to maintain generalization across depths. Implemented with XGBRegressor from \cite{XGBoost}.

\textbf{Cat Boost (CAT)} \\
A gradient boosting algorithm that is particularly designed to handle categorical features effectively, using techniques like ordered boosting and target statistics to reduce overfitting. CAT is included as a complementary approach to assess whether its regularization mechanisms and ordered boosting scheme improve generalization. Implemented with CatBoostRegressor \cite{CatBoost}.

\textbf{Multilayer Perceptron Network (MLP)} \\
A feed-forward artificial neural network with one or more hidden layers. Each layer consists of neurons applying linear transformations followed by non-linear activation functions. The network is trained via backpropagation to minimize a loss function \cite{MLPRegressor}. The MLP is included with the intention of approximating highly nonlinear reflectance–chlorophyll relationships that may not be captured by tree-based or linear models. Implemented with MLPRegressor from \texttt{sklearn}.

\textbf{Ensemble (ENS)} \\
A meta-model that combines predictions from multiple base models (e.g. averaging, stacking) in order to improve predictive performance and robustness, by leveraging their complementary strengths and reducing individual model’s error variance. In this instance, the ensemble is built by averaging the other models.

\subsection{Buoy data}
\begin{figure}
    \centering
    \includegraphics[width=0.46\textwidth]{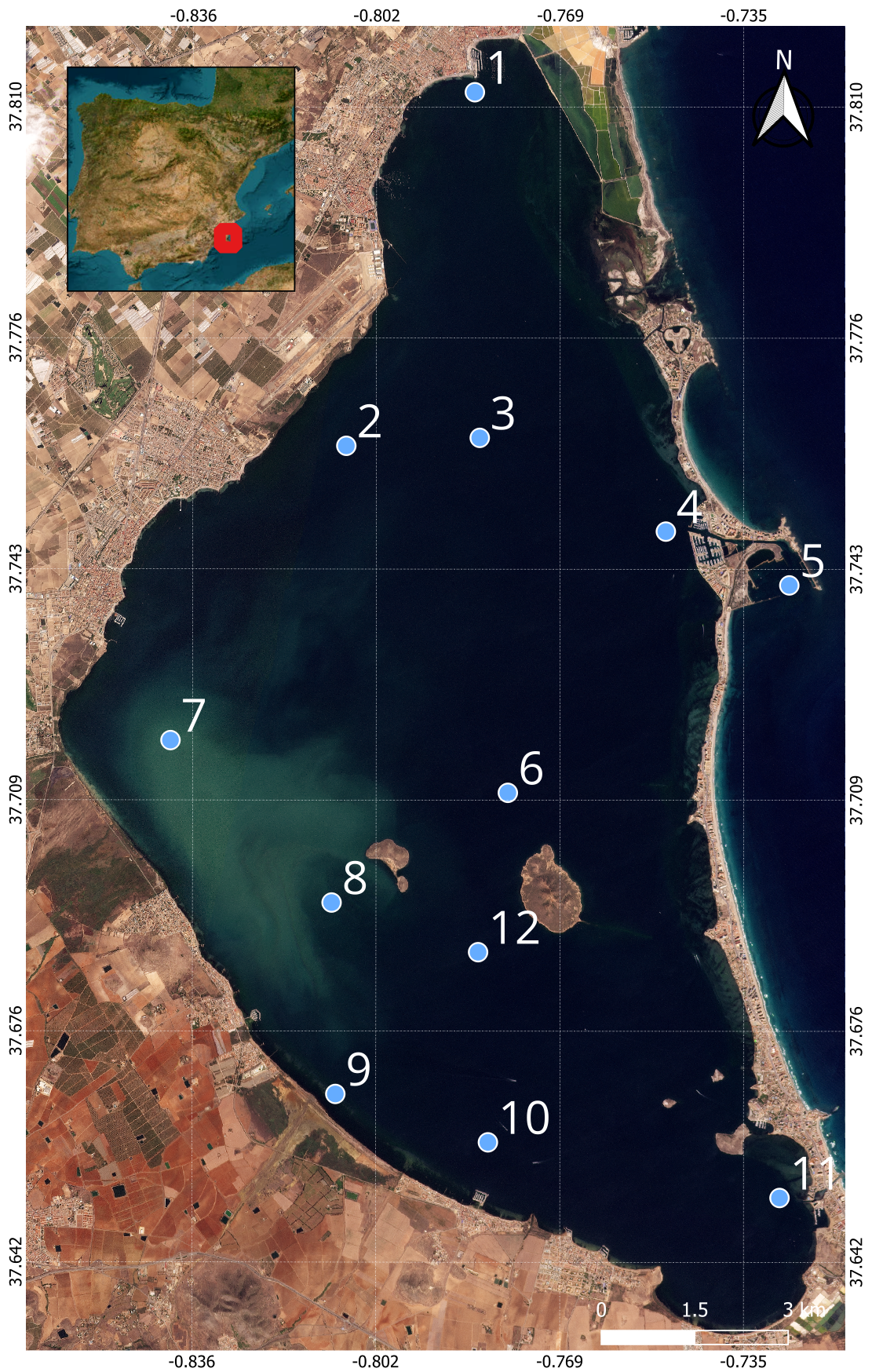}
    \caption{Buoy distribution in the Mar Menor. Top left map of Iberian Peninsula extracted from \cite{ORSMaps}}
    \label{fig:mapaboyas}
\end{figure}

There is a set of 12 buoys located in the Mar Menor, shown in Figure \ref{fig:mapaboyas} and located in the coordinates from Table \ref{tab:buoycoordinates} according to \cite{tomadedatosenelmar} . These buoys measure several water quality parameters, such as oxygen, salinity, Chl-a, temperature or turbidity. 

\begin{table}[!b]
    \caption{Control Points with Latitude and Longitude}
    \label{tab:buoycoordinates}
    \centering
    \begin{tabular}{ccc}
        \hline
        Buoy & Latitude   & Longitude  \\
        \hline
        CTD-1           & 37.811800 & -0.784483 \\
        CTD-2           & 37.760617 & -0.807800   \\
        CTD-3           & 37.761783 & -0.783550  \\
        CTD-4           & 37.748233 & -0.749617 \\
        CTD-5           & 37.740450 & -0.727117 \\
        CTD-6           & 37.710417 & -0.773833 \\
        CTD-7           & 37.718000 & -0.839783 \\
        CTD-8           & 37.694517 & -0.810400 \\
        CTD-9           & 37.666817 & -0.809683 \\
        CTD-10          & 37.659833 & -0.781967 \\
        CTD-11          & 37.651800 & -0.728883 \\
        CTD-12          & 37.68735  & -0.783783 \\
        \hline
    \end{tabular}
\end{table}

The data gathered comes from two complementary sources \cite{universidadpolitecnicadecartagenaDatosBoyasUPCT} and \cite{institutomurcianodeinvestigacionydesarrolloagrarioymedioambientalDatosBoyasIMIDA}. The latter is no longer available due to the end of the project related to that website. For simplicity, the initial data source will be referred to as Polytechnic University of Cartagena (UPCT), and the subsequent one as Murcian Institute of Agricultural and Environmental Research and Development (IMIDA). It is hypothesized that both sources are located in the same locations; however, the metadata and information in the data sources do not specify if the buoys are the same. 
The UPCT dataset encompasses the period from 2017-05-19 to 2024-07-18. Buoys measure at increasing depths, ranging from 0.5 meters to 5.0 meters. In areas where it is feasible to descend to these depths, measurements are typically taken in steps of 0.5 meters.
The IMIDA dataset ranges form 2016-06-29 to 2023-08-24 and measurements are taken at surface level (0.0 meters) and then at 1, 2, 3, 4 and 5 meters.
In both cases, data is retrieved approximately once a week, despite variations over time and across seasons.

An analysis comparing both datasets at each depth reveals that, in most cases, the differences between the measurements are not significant. Therefore, the merged data goes from 2016-06-29 to 2024-07-18, achieving close to a decade-long dataset.

The CTD sensors on the buoys continuously measure three parameters: conductivity, temperature, and depth. CTDs are typically equipped with additional sensors that facilitate the recording of a broader range of biochemical variables. In this study, Chl-a measurement is performed using fluorometers, taking advantage of the red autofluorescence of this molecule when exposed to blue light. The measurement of Chl-a \cite{tomadedatosenelmar} is expressed in milligrams per cubic meter (mg/m$^3$), which is equivalent to micrograms per liter ($\mu$g/L). The measurement of this parameter by fluorometry may exhibit slight discrepancies compared to the measurement obtained by spectrophotometry on an extract of this pigment due to non-photochemical quenching or varying fluorescence yields\cite{parsonsManualChemicalBiological1984}. It should also be noted that in shallow waters such as the Mar Menor, traces of submerged vegetation that keep Chl-a or active pheopigments can be found in the water column. The methodology proposed prioritizes the high temporal resolution and continuous monitoring provided by buoy networks. The cross-consistency observed between the UPCT and IMIDA datasets, combined with the use of gradient boosting models known for their robustness to label noise, ensures that the predicted maps remain representative of the lagoon’s ecological state.

\subsection{Satellite imagery}\label{subsec:satelliteimagery}

Sentinel 2 imagery is accessible to the public via the Copernicus Browser \cite{CopernicusBrowser}. Within this interface, users can manually define a bounding box to delineate the area of interest, specify a date range, select the desired level of processing, choose the data source and download SAFE files. These SAFE files correspond to a large area, particularly a tile, and contains an image file for each spectral band, along with metadata. A different approach that is considerably more suitable for the downloading of multiple files is through the Copernicus API, a method that permits the automation of the process. In our case, we use a Python script to achieve this objective.

This script was developed to access the Copernicus Data Space Ecosystem repository through the boto3 \cite{Boto314026Documentation} interface, The dates of interest and the Sentinel 2 tile identifier corresponding to the study area (Tile ID: 30SXG) were specified. The script dynamically constructs the search prefixes based on the date, following the hierarchical structure of the repository. The specific designation for this satellite is "Sentinel 2/MSI/L1C\_N0500/$<$year$>$/$<$month$>$/$<$day$>$" or "Sentinel 2/MSI/L1C/$<$year$>$/$<$month$>$/$<$day$>$", depending on the date. For each shot, the product in SAFE format containing the required Tile ID is selected, and all spectral bands are downloaded together with the associated metadata files to a local directory. By downloading L1C products as files in SAFE format, we can process them later on with SNAP.

The initial goal was to maximize the number of satellite images available for analysis, starting with dates in which \textit{in situ} Chl-a measurements from the buoys coincided with a Sentinel 2 acquisition. Subsequent analysis revealed that approximately 50\% of the images were predominantly black and had a file size of approximately 40 MB, which is significantly smaller than the typical range of 600 to 800 MB for standard images. The images in question are removed from the collection, and then we proceed to filter images based on cloud coverage. This step was divided into two parts: first, a script automatically read the Sentinel Catalog and requested cloud coverage data in the area of interest; then, the images were filtered by visual inspection to ensure that there were no significant clouds, particularly in the locations where the buoys were positioned. After thoroughly reviewing the 180 images downloaded from the selection, only 41 were deemed useful.

\section{Methodology}\label{sec:methodology}

This section describes the processing applied to buoy and satellite data separately, the characteristics of the posterior merged data, the band combinations introduce as additional features and the datasets, understood as the set of target chlorophyll depth, the processing method, and the reflectance aggregation window.

\subsection{Buoy Data preprocessing}

Once the buoy data is downloaded from the two sources, UPCT and IMIDA, the NC (NetCDT) file is converted to a CSV for each buoy. These CSVs have a column for the date of the measurement, and then a column for each depth. Since there are two buoy data sources the objective is to merge them. The process for the merge consists on the following:
\begin{itemize}
    \item Load the chlorophyll CSVs along with another CSV that contains the buoys' locations and unify the name of the buoys and columns from both sources, since their default names differs.
    \item Use a function to average Chl-a in a set of given depths. The groups are 0 to 1, 1 to 2, 2 to 3, and 3 to 4 meters in depth. Those Chl-a levers are the ones which will be predicted.
    \item Use another function to combine the averaged Chl-a, dates, identifiers and locations.
    \item Save a CSV for each depth from both UPCT and IMIDA.
    \item Load the pair of CSVs for each depth and: if there is only a measurement for a date, that one is selected, whereas if both CSV contain a measurement for the same date, we take the mean value.
    \item Save the final CSV with the columns ``Date'', ``Buoy'' and ``Chl'' for each depth.
\end{itemize}

\subsection{Satellite data processing}

The SAFE files are processed with SNAP. In order to apply the atmospheric correction algorithms from the C2RCC project the following process is performed:
\begin{enumerate}
    \item Resample the image to 10 m with the ``Nearest'' method. The decision of resampling to 10 m rather than downsampling to the native 60 m of certain bands, was driven by the need to maintain high feature granularity. A priori, it is not possible to determine if aggregating spectral information into larger pixels (e.g., 60 m) would improve model performance or, conversely, act as a low-pass filter that obscures spatial variability. By maintaining a 10 m base resolution, the ML models are provided with the highest spatial resolution available, allowing the subsequent window aggregation strategies (from 3$\times$3 to 15$\times$15 windows) to systematically explore the optimal scale for noise reduction without sacrificing the underlying spatial structure of the data.
    \item Crop the subset of the image with the geocoordinates 37.82, -0.867, 37.62, -0.7 for north latitude, west longitude, south latitude, and east latitude bounds, respectively, to delineate the Mar Menor lagoon area for targeted analysis.
    \item Apply de C2RCC processing for S2-SMI. The outputs selected were TOA reflectantes (uncorrected), normalized water-leaving reflectances (named rhown) and atmospherically corrected angular dependent reflectances (rhow) \cite{SNAPDataProcessors}. While Remote Sensing Reflectance ($R_{rs}$) is also commonly reported in literature, rhow and $R_{rs}$ are linearly related by a factor of $\pi$\cite{petersStepForumC2RCC2020, petersStepForumC2RCC2018}, and are more common when dealing with shallow waters \cite{katlaneRemoteSensingTurbidity2024, pereira-sandovalEvaluationAtmosphericCorrection2019}. Therefore, the choice of water-leaving reflectances over $R_{rs}$ does not affect the predictive power or the internal logic of the ML algorithms, as these models are invariant to linear scaling of the input features.
    \item Take the subset of bands to leave out residual metadata and obtain a TIFF file with 28 bands: B1 to B12 from TOA reflectances, B1 to B8A from rhow, B1 to B6 from rhown, and a c2rcc\_flags necessary to export the product.
\end{enumerate}

Once the flux is defined, it is possible to automate the process with the SNAP functionality Graph Processing Framework \cite{SNAPGraphProcessing}. This framework is used with XML files that define the processing workflow described above, which is further customized through a properties file to manage the parameters of each step. To fully automate the process, a bash script is then employed. The script iterates through all the dates of interest, applies the workflow specified in the XML file, and saves the resulting outputs as TIFF files. There are three iterations for each date to obtain the image processed with C2RCC, C2X and C2X-Complex. 

The TIFF files were processed using Python with the aim of aggregating pixels within windows of different sizes to assess which aggregation scale yields the most reliable results. For each of the C2RCC neural networks, five window sizes were tested: 1$\times$1, 3$\times$3, 5$\times$5, 9$\times$9, and 15$\times$15 pixels. In practice, this means that reflectance values at the buoy locations were derived by calculating the mean of the surrounding pixels rather than relying solely on the exact pixel corresponding to the buoy position. This approach was adopted because the processed images often exhibited substantial pixel-to-pixel variation even within very small areas. By averaging across a window, local fluctuations are smoothed, providing a more representative estimate of the reflectance at a given location. The outputs of this procedure were stored as CSV files, with one file generated for each combination of neural network and window size.

\subsubsection{Band combinations}

Following the ideas from related works previously mentioned, along with more options provided by \cite{delegidoModeloEmpiricoPara2014}, many band combinations were explored as inputs for the models. In preliminary tests, formulas from \cite{yeleyMarMenorLagoon2022}, \cite{erenaMonitoringCoastalLagoon2019} and \cite{gomez-jakobsenMonitoringChlorophyllConcentration2025} where considered, but were finally discarded due to not improving the models. Therefore, the combinations used are:

\begin{itemize}
    \item Normalized difference $$\frac{R(\lambda_1) - R(\lambda_2)}{R(\lambda_1) + R(\lambda_2)}$$
    \item Dall-Gitelson $$\left(\frac{1}{R(\lambda_1)}-\frac{1}{R(\lambda_2)}\right) \times R(\lambda_3)$$
    \item Four-band normalized difference index $$\frac{R(\lambda_1) - R(\lambda_2)}{R(\lambda_3) + R(\lambda_4)}$$
    \item Difference of inverses $$\frac{1}{R(\lambda_1)}-\frac{1}{R(\lambda_2)}$$
    \item Four-band ratio difference $$\frac{R(\lambda_1)}{R(\lambda_2)}-\frac{R(\lambda_3)}{R(\lambda_4)}$$
    \item Three-band normalized sum $$\frac{R(\lambda_i) + R(\lambda_{i+2})}{R(\lambda_i) + R(\lambda_{i+1})}$$ where $\lambda_{i+2} > \lambda_{i+1} > \lambda_i$.
\end{itemize}

These formulas capture most of the band combinations explored in the majority of the related literature, with the exception of empirical methods and manually adjusted expressions.
When applying all of these, use symmetrical terms were avoided, i.e., those that, when two bands are swapped, give the same result, both with the same sign and position, since the correlation with the target will be the same, as well as those that produce the same result but with opposite sign.

\subsection{Datasets}

With the buoy and satellite data prepared, the next step is combine them into merged datasets, and obtain a final dataset for each group of depth, processing, and window aggregation. Additionally, C2RCC, C2X and C2X-Complex add another layer of dimensionality because they derived into two versions: atmospherically corrected and normalized reflectances, named rhow y rhown, respectively; whereas TOA reflectances are obtained from one of the previous, and are a single set of bands since they have not undergone any processing.
The merging of buoy and satellite data is performed exclusively when both measurement sources are collected on the same date. The proposal in \cite{gomezNewApproachMonitor2021} to include more observations by using satellite images taken one or two days before or after the chlorophyll measurement was considered. However, since chlorophyll is a highly variable parameter, the data matching time window was restricted to a same-day criterion to minimize variance. Sentinel 2 acquisitions over Mar Menor typically occurred between 10:30 and 11:30 \cite{CopernicusBrowser}, and buoy measurements are usually taken between 10:00 and 13:00. Consequently, the maximum temporal offset between the two sensing platforms is approximately 2.5 hours. Given the hydrodynamic stability of the lagoon under the selected cloud-free conditions, Chl-a concentrations are not expected to undergo significant biological or spatial fluctuations within this narrow window. This synchronization ensures that the spectral signatures captured by the satellite are representative of the physical samples measured by the in situ sensors, thereby providing a high-fidelity dataset for the training and validation of the ML architectures.
At this point, an auxiliary high-chlorophyll class was defined as the 90th percentile of the Chl-a concentration to ensure a balanced representation of scarce high-concentration samples, as well as a categorical variable to include the season of the observation.

The preparation of the datasets enhances the statistical robustness of the models by ensuring that buoy and satellite measurements are properly aligned, and by using aggregation windows to smooth local variability in reflectance values. In this way, the resulting datasets are more stable and less affected by noise, which supports more reliable training and better generalization.

\begin{figure}
    \centering
    \includegraphics[width=0.48\textwidth]{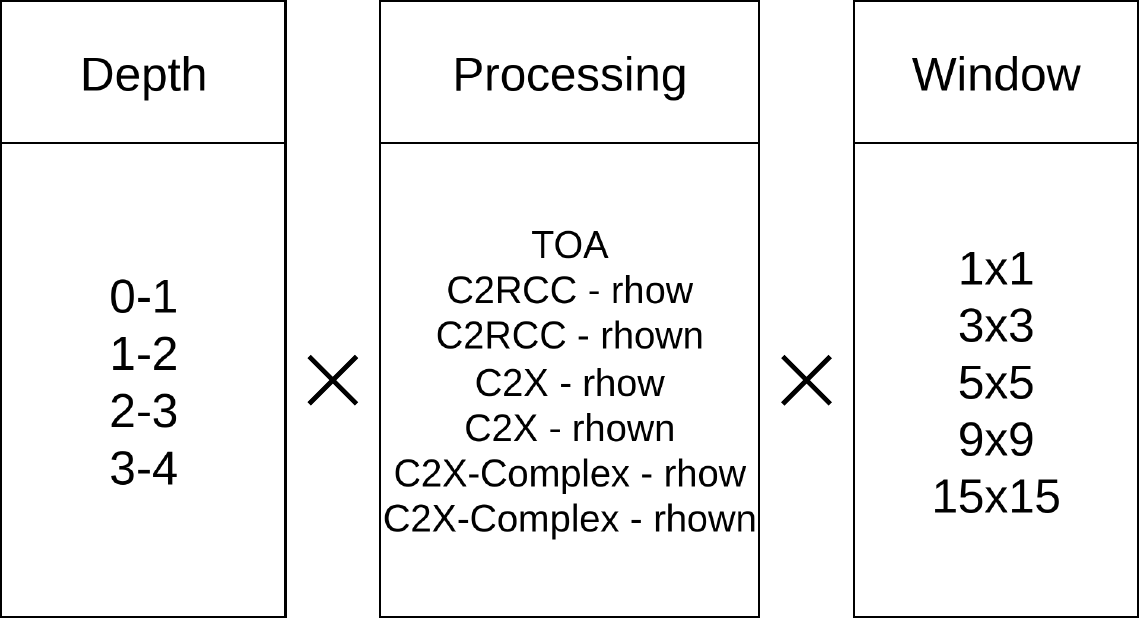}
    \caption{Dataset combinations. All depths (in meters), processing methods, and aggregation windows (in pixels) were combined combined in every possible way.} 
    \label{fig:DatasetCombinations}
\end{figure}

Figure \ref{fig:DatasetCombinations} shows more clearly the dimensionality of the problem and what are the options explored. This is done with the aim of identifying, for each depth, what type of processing and window aggregation allows for a more accurate prediction of Chl-a.

\section{Experiments}\label{sec:exp}

The experiments conducted are described in this section. The scripts are available in the Github repository \url{https://github.com/Antonio-MI/mar-menor-chl}. The experiments were run on four virtual machines with Ubuntu 24.04 LTS, 128 GB of RAM, and an AMD EPYC 9554 64-Core processor each. The code was implemented using Python 3.11.0, and SNAP processing with SNAP 12.0.0.

The procedure followed to train the models consists on an iterative process in which there is a preliminary training, an hyperparameter selection phase, and a final training and evaluation.

The preliminary training stage began with setting the model parameters to their default values, removing missing entries on the datasets, and training with a 5-fold stratified cross-validation scheme. Using this cross-validation setup (described in more detail in the final training phase), the top ten performing datasets were selected for each processing method. This was based only on the highest R$^2$ and Root Mean Squared Logarithmic Error (RMSLE) values. This means that, across all the algorithms, the best one determines whether a dataset is selected or not.

After preliminary training, hyperparameter optimization was performed using the Optuna framework \cite{akiba2019optuna}, which allows to create a ``study'' instance to optimize an objective function for a particular value specified by the user. Given the high variability of chlorophyll-a concentrations in the Mar Menor, which range from oligotrophic levels to extreme eutrophic peaks, a multi-objective optimization was implemented. The study was configured to simultaneously minimize the Root Mean Squared Logarithmic Error (RMSLE) and maximize the $R^2$ score. During this phase, a 4-fold stratified (with the high-chlorophyll class) cross-validation scheme was employed to evaluate those two metrics and tune the models' hyperparameters. The inclusion of RMSLE as one of the objectives is crucial for this context, as it provides a more balanced penalization of relative errors across different orders of magnitude. On the other hand, the $R^2$ accounts for global variance explanation. This process led to the identification of the Pareto front, which consists of the set of optimal solutions where it is impossible to improve one metric without further degrading the other. By exploring this front, Optuna identified hyperparameter configurations that provide a balanced trade-off between overall predictive power and stability across the entire concentration spectrum of the water column.

Given the high dimensionality of the experimental design, which involved 100 independent seeds, multiple water column depths, and various dataset configurations, the total number of required training trials was exceptionally high. To manage this computational load and maximize the system's efficiency, a parallel execution strategy was implemented. This approach involved parallelizing the Optuna optimization process alongside the execution of individual ML models that support multi-threading.

Once the hyperparameters had been optimized for each dataset--defined by the target Chl-a depth, the processing method, and the reflectance aggregation window--the final training phase was carried out for each depth. Previous iterations of the study revealed that models trained on a single seed could achieve high performance; however, these results exhibited significant sensitivity to the initial data partition, leading to inconsistent estimates of the model's actual precision. To address this and ensure the statistical robustness of the findings, the entire experimental process was repeated across 100 different random seeds. In the proposed workflow, for each seed, the data were partitioned into a training set (80\%) and a test set (20\%). This split was stratified using the auxiliary high-chlorophyll class.

Input variables ($X$) were separated from the target ($y$), strictly excluding the high chlorophyll feature from the input to prevent data leakage. For distance-based models (such as KNN or SVR), data were scaled using a RobustScaler before training, and predictions were subsequently back-transformed. In contrast, tree-based models were trained directly on the raw features. The ensemble model was constructed by calculating the simple arithmetic mean of the predictions generated by all individual base models. 

Finally, performance metrics ($R^2$, RMSLE, and RMSE) were computed for both training and test sets. All results, including best parameters and predictions, were serialized using Pickle to ensure full reproducibility of the experiment.
After evaluating all the experiments with different seeds, the results were aggregated to obtain a mean value and its respective standard deviation. Those results are discussed in the following section.

\subsection{Results}

Model performance was evaluated using three complementary metrics: the root mean squared logarithmic error (RMSLE), the coefficient of determination ($R^2$) and the root mean squared error (RMSE). RMSLE was selected because Chl-a data gathered in this study follows a log-normal distribution spanning several orders of magnitude, and therefore the error should be optimized in the logarithmic space. Optimizating this metric would lead to stable predictions if for both low and high Chl-a scenarios.
R-Squared, or $R^2$, quantifies the proportion of variance in the observed values that is explained by the model, with values closer to 1 indicating a better fit. RMSE measures the average magnitude of the prediction errors, expressed in the same units as the target variable, and lower values denote higher accuracy. Although this metric was not prioritized during model optimization, it is included to maintain comparability with related works.

Formally, given observed values $y_i$, predicted values $\hat{y}_i$, and their mean $\bar{y}$, the metrics are defined as:

\begin{equation}
    R^2 = 1 - \frac{\sum_{i=1}^{n} (y_i - \hat{y}_i)^2}{\sum_{i=1}^{n} (y_i - \bar{y})^2}   
\end{equation}

\begin{equation}
    RMSLE = \sqrt{\frac{1}{n}\sum_{i=1}^{n} (\log(1 + y_i) - \log(1 + \hat{y}_i))^2}
\end{equation}

\begin{equation}
    RMSE = \sqrt{\frac{1}{n}\sum_{i=1}^{n} (y_i - \hat{y}_i)^2}
\end{equation}

Since models and datasets are grouped for each depth, result are also shown by depth. To ensure conciseness, we only report the test performance metrics of RMSLE and $R^2$, but complete test and train metrics, including RMSE as well, are also available in the Supplementary Material. 

Metrics shown in the following tables provide the most relevant assessment of the models’ predictive ability. Each table consists of ten rows, one for each of the datasets, determined by type of processing, window aggregation and depth, and a column for each model. Entries on the tables are ordered taking the best value (lower RMSLE or higher R$^2$) of each row, across all the models. The best model for each scenario is highlighted in bold\footnote{Linear Regression, ElasticNet and Random Forest were also evaluated, but are omitted here to maintain a legible table while keeping the best-scoring models. Full results are available in the Supplementary Material.}. While a robust $R^2$ is expected to approach unity, acceptable RMSLE values are inherently context-dependent. In this study, based on the analytical relationship between log-error and variance explanation during training, an RMSLE range between 0.35 and 0.4 is categorized as satisfactory, as it provides a reliable estimation for general monitoring purposes. Consequently, values below 0.35 are considered closer to optimal, whereas values exceeding 0.4 are deemed insufficient for this particular application.

Tables \ref{tab:test_RMSLE_0-1} and \ref{tab:test_R2_0-1} display the results for the most superficial layer. CatBoost was the best-performing model, obtaining the best metrics for both RMSLE and $R^2$, while KNN and SVR also performed consistently. Among datasets, C2RCC and C2X-Complex with and without normalized reflectances, as well as TOA, yielded similar metrics. Likewise, none of the window aggregations performed significantly better than the others, except for the configurations employing 1$\times$1 or 3$\times$3 windows, which did not make the top ten results. RMSLE for the best configurations is below 0.34 and $R^2$ above 0.75, making predictions for the most superficial depth quite reliable for monitoring.

\begin{table*}[!t]
    \caption{Test RMSLE for depth 0-1 meters. The best model for each scenario is highlighted in bold.}
    \label{tab:test_RMSLE_0-1}
    \centering
    \resizebox{\textwidth}{!}{%
    \begin{tabular}{lcccccccccc}
        \hline
        \textbf{Model} & \textbf{CAT} & \textbf{ENS} & \textbf{KNN} & \textbf{LBM} & \textbf{MLP} & \textbf{SVR} & \textbf{XGB} \\
        \hline
        TOA\_15x15\_depth\_in\_0\_1                & 0.337 ± 0.047          & 0.362 ± 0.046 & \textbf{0.331 ± 0.049} & 0.385 ± 0.064 & 0.409 ± 0.073 & 0.426 ± 0.038          & 0.377 ± 0.061 \\
        C2X-Complex\_rhow\_15x15\_depth\_in\_0\_1  & \textbf{0.337 ± 0.036} & 0.356 ± 0.034 & 0.367 ± 0.042          & 0.363 ± 0.043 & 0.392 ± 0.048 & 0.354 ± 0.033          & 0.353 ± 0.041 \\
        C2RCC\_rhow\_5x5\_depth\_in\_0\_1          & \textbf{0.346 ± 0.035} & 0.352 ± 0.035 & 0.356 ± 0.044          & 0.375 ± 0.039 & 0.388 ± 0.051 & 0.368 ± 0.037          & 0.366 ± 0.039 \\
        C2RCC\_rhown\_5x5\_depth\_in\_0\_1         & \textbf{0.347 ± 0.036} & 0.349 ± 0.034 & 0.356 ± 0.045          & 0.363 ± 0.038 & 0.381 ± 0.047 & 0.365 ± 0.036          & 0.355 ± 0.038 \\
        C2X-Complex\_rhow\_9x9\_depth\_in\_0\_1    & \textbf{0.349 ± 0.042} & 0.370 ± 0.042 & 0.396 ± 0.054          & 0.378 ± 0.048 & 0.421 ± 0.053 & 0.360 ± 0.039          & 0.363 ± 0.042 \\
        C2X-Complex\_rhow\_5x5\_depth\_in\_0\_1    & 0.364 ± 0.043          & 0.381 ± 0.048 & 0.389 ± 0.056          & 0.409 ± 0.051 & 0.436 ± 0.050 & \textbf{0.356 ± 0.035} & 0.396 ± 0.045 \\
        C2X-Complex\_rhown\_15x15\_depth\_in\_0\_1 & 0.359 ± 0.046          & 0.366 ± 0.041 & 0.374 ± 0.048          & 0.383 ± 0.052 & 0.395 ± 0.043 & \textbf{0.358 ± 0.033} & 0.377 ± 0.046 \\
        C2X-Complex\_rhown\_9x9\_depth\_in\_0\_1   & \textbf{0.362 ± 0.042} & 0.377 ± 0.046 & 0.401 ± 0.058          & 0.386 ± 0.051 & 0.408 ± 0.058 & 0.365 ± 0.040          & 0.382 ± 0.042 \\
        C2RCC\_rhow\_15x15\_depth\_in\_0\_1        & \textbf{0.367 ± 0.048} & 0.375 ± 0.043 & 0.378 ± 0.062          & 0.395 ± 0.050 & 0.416 ± 0.060 & 0.385 ± 0.045          & 0.386 ± 0.048 \\
        C2RCC\_rhow\_9x9\_depth\_in\_0\_1          & \textbf{0.369 ± 0.041} & 0.382 ± 0.041 & 0.384 ± 0.052          & 0.405 ± 0.045 & 0.421 ± 0.055 & 0.382 ± 0.042          & 0.391 ± 0.043 \\
        \hline
    \end{tabular}
    }
\end{table*}

\begin{table*}[!t]
    \caption{Test $R^2$ for depth 0-1 meters. The best model for each scenario is highlighted in bold.}
    \label{tab:test_R2_0-1}
    \centering
    \resizebox{\textwidth}{!}{%
    \begin{tabular}{lcccccccccc}
        \hline
        \textbf{Model} & \textbf{CAT} & \textbf{ENS} & \textbf{KNN} & \textbf{LBM} & \textbf{MLP} & \textbf{SVR} & \textbf{XGB} \\
        \hline
        C2RCC\_rhow\_5x5\_depth\_in\_0\_1          & \textbf{0.767 ± 0.082} & 0.768 ± 0.072          & 0.756 ± 0.099          & 0.707 ± 0.118 & 0.743 ± 0.138 & 0.706 ± 0.086 & 0.734 ± 0.100 \\
        C2RCC\_rhown\_5x5\_depth\_in\_0\_1         & 0.763 ± 0.083          & \textbf{0.767 ± 0.070} & 0.758 ± 0.093          & 0.719 ± 0.123 & 0.751 ± 0.113 & 0.701 ± 0.078 & 0.742 ± 0.106 \\
        TOA\_15x15\_depth\_in\_0\_1                & 0.743 ± 0.117          & 0.725 ± 0.092          & \textbf{0.764 ± 0.131} & 0.626 ± 0.201 & 0.659 ± 0.209 & 0.411 ± 0.091 & 0.618 ± 0.195 \\
        C2X-Complex\_rhow\_9x9\_depth\_in\_0\_1    & \textbf{0.763 ± 0.082} & 0.736 ± 0.083          & 0.670 ± 0.122          & 0.708 ± 0.122 & 0.656 ± 0.139 & 0.699 ± 0.098 & 0.732 ± 0.102 \\
        C2X-Complex\_rhow\_15x15\_depth\_in\_0\_1  & \textbf{0.755 ± 0.098} & 0.729 ± 0.086          & 0.711 ± 0.110          & 0.662 ± 0.143 & 0.704 ± 0.112 & 0.698 ± 0.111 & 0.696 ± 0.118 \\
        C2X-Complex\_rhow\_5x5\_depth\_in\_0\_1    & \textbf{0.737 ± 0.089} & 0.709 ± 0.089          & 0.666 ± 0.126          & 0.639 ± 0.143 & 0.647 ± 0.131 & 0.693 ± 0.107 & 0.640 ± 0.126 \\
        C2RCC\_rhow\_15x15\_depth\_in\_0\_1        & 0.730 ± 0.098          & 0.723 ± 0.089          & \textbf{0.733 ± 0.148} & 0.669 ± 0.132 & 0.682 ± 0.147 & 0.677 ± 0.097 & 0.679 ± 0.131 \\
        C2X-Complex\_rhown\_9x9\_depth\_in\_0\_1   & \textbf{0.733 ± 0.092} & 0.716 ± 0.088          & 0.671 ± 0.123          & 0.667 ± 0.128 & 0.673 ± 0.141 & 0.689 ± 0.096 & 0.674 ± 0.109 \\
        C2RCC\_rhow\_9x9\_depth\_in\_0\_1          & \textbf{0.726 ± 0.098} & 0.721 ± 0.086          & 0.724 ± 0.127          & 0.658 ± 0.136 & 0.678 ± 0.158 & 0.691 ± 0.092 & 0.691 ± 0.119 \\
        C2X-Complex\_rhown\_15x15\_depth\_in\_0\_1 & \textbf{0.709 ± 0.115} & 0.705 ± 0.100          & 0.702 ± 0.125          & 0.639 ± 0.178 & 0.686 ± 0.113 & 0.691 ± 0.103 & 0.640 ± 0.138 \\
        \hline
    \end{tabular}
    }
\end{table*}

The next segment of the water column is between 1 and 2 meters deep. The results for this segment are in Tables \ref{tab:test_RMSLE_1-2} and \ref{tab:test_R2_1-2}. In this case, the datasets that were processed with C2X-Complex predominated among the best results. The top performance was achieved with ``C2X-Complex\_rhow\_15x15'', followed by the same configuration with a 9$\times$9 window. While SVR also provided the strong results, CatBoost consistently outperformed other models in most scenarios. The metrics for this depth remain robust, with an $R^2$ above 0.75 and an RMSLE ranging between 0.35 and 0.4. This performance is interpreted as satisfactory, showing only a marginal increase in error compared to the superficial layer.

When comparing the performance obtained in the upper layers with the studies summarized in Table \ref{tab:related} our metrics demonstrate a high degree of competitiveness with previous findings for the Mar Menor. However, it is worth noting that our dataset spans nearly a decade, covers the widest range of Chl-a concentrations reported so far, offers the highest spatial resolution, and includes a complete validation framework that takes into consideration the inherent variability of the dataset by repeting the experiments with 100 different seeds. These factors make the results particularly robust, albeit at the expense of a more complex and computationally demanding processing workflow.

\begin{table*}[!t]
    \caption{Test RMSLE for depth 1-2 meters. The best model for each scenario is highlighted in bold.}
    \label{tab:test_RMSLE_1-2}
    \centering
    \resizebox{\textwidth}{!}{%
    \begin{tabular}{lcccccccccc}
        \hline
        \textbf{Model} & \textbf{CAT} & \textbf{ENS} & \textbf{KNN} & \textbf{LBM} & \textbf{MLP} & \textbf{SVR} & \textbf{XGB} \\
        \hline
    C2X-Complex\_rhow\_15x15\_depth\_in\_1\_2 & \textbf{0.365 ± 0.048} & 0.385 ± 0.048 & 0.392 ± 0.055 & 0.394 ± 0.053 & 0.429 ± 0.049 & 0.375 ± 0.042          & 0.380 ± 0.050 \\
    C2X-Complex\_rhow\_9x9\_depth\_in\_1\_2   & \textbf{0.382 ± 0.047} & 0.398 ± 0.051 & 0.419 ± 0.059 & 0.393 ± 0.053 & 0.454 ± 0.065 & 0.387 ± 0.047          & 0.389 ± 0.048 \\
    C2X-Complex\_rhown\_9x9\_depth\_in\_1\_2  & \textbf{0.392 ± 0.047} & 0.403 ± 0.052 & 0.425 ± 0.064 & 0.412 ± 0.048 & 0.442 ± 0.064 & 0.395 ± 0.049          & 0.406 ± 0.043 \\
    C2X-Complex\_rhow\_5x5\_depth\_in\_1\_2   & 0.395 ± 0.047          & 0.409 ± 0.052 & 0.418 ± 0.054 & 0.426 ± 0.050 & 0.463 ± 0.060 & \textbf{0.393 ± 0.038} & 0.412 ± 0.048 \\
    C2X-Complex\_rhow\_3x3\_depth\_in\_1\_2   & \textbf{0.394 ± 0.042} & 0.412 ± 0.049 & 0.420 ± 0.051 & 0.422 ± 0.051 & 0.466 ± 0.050 & 0.409 ± 0.035          & 0.416 ± 0.046 \\
    C2X-Complex\_rhown\_3x3\_depth\_in\_1\_2  & \textbf{0.395 ± 0.045} & 0.412 ± 0.051 & 0.426 ± 0.051 & 0.419 ± 0.058 & 0.454 ± 0.055 & 0.413 ± 0.038          & 0.410 ± 0.052 \\
    C2RCC\_rhown\_3x3\_depth\_in\_1\_2        & \textbf{0.400 ± 0.041} & 0.405 ± 0.043 & 0.404 ± 0.046 & 0.425 ± 0.046 & 0.423 ± 0.049 & 0.408 ± 0.042          & 0.419 ± 0.044 \\
    C2X-Complex\_rhown\_5x5\_depth\_in\_1\_2  & 0.402 ± 0.045          & 0.413 ± 0.051 & 0.424 ± 0.053 & 0.432 ± 0.051 & 0.450 ± 0.047 & \textbf{0.401 ± 0.042} & 0.421 ± 0.048 \\
    C2X\_rhow\_5x5\_depth\_in\_1\_2           & 0.407 ± 0.041          & 0.418 ± 0.042 & 0.422 ± 0.040 & 0.427 ± 0.052 & 0.462 ± 0.056 & \textbf{0.406 ± 0.035} & 0.424 ± 0.046 \\
    C2X\_rhow\_3x3\_depth\_in\_1\_2           & 0.422 ± 0.041          & 0.427 ± 0.044 & 0.433 ± 0.045 & 0.442 ± 0.050 & 0.471 ± 0.052 & \textbf{0.421 ± 0.036} & 0.440 ± 0.047 \\
        \hline
    \end{tabular}
    }
\end{table*}

\begin{table*}[!t]
    \caption{Test $R^2$ for depth 1-2 meters. The best model for each scenario is highlighted in bold.}
    \label{tab:test_R2_1-2}
    \centering
    \resizebox{\textwidth}{!}{%
    \begin{tabular}{lcccccccccc}
        \hline
        \textbf{Model} & \textbf{CAT} & \textbf{ENS} & \textbf{KNN} & \textbf{LBM} & \textbf{MLP} & \textbf{SVR} & \textbf{XGB} \\
        \hline
    C2X-Complex\_rhow\_15x15\_depth\_in\_1\_2 & \textbf{0.759 ± 0.097} & 0.729 ± 0.111 & 0.714 ± 0.142 & 0.670 ± 0.190 & 0.684 ± 0.128 & 0.728 ± 0.104          & 0.715 ± 0.123 \\
    C2X-Complex\_rhow\_9x9\_depth\_in\_1\_2   & \textbf{0.756 ± 0.104} & 0.723 ± 0.116 & 0.673 ± 0.148 & 0.715 ± 0.148 & 0.613 ± 0.202 & 0.740 ± 0.085          & 0.728 ± 0.126 \\
    C2X-Complex\_rhow\_5x5\_depth\_in\_1\_2   & \textbf{0.748 ± 0.096} & 0.711 ± 0.130 & 0.664 ± 0.173 & 0.669 ± 0.166 & 0.625 ± 0.173 & 0.747 ± 0.085          & 0.696 ± 0.126 \\
    C2RCC\_rhown\_3x3\_depth\_in\_1\_2        & \textbf{0.743 ± 0.115} & 0.730 ± 0.113 & 0.732 ± 0.129 & 0.666 ± 0.181 & 0.727 ± 0.131 & 0.722 ± 0.115          & 0.681 ± 0.158 \\
    C2X-Complex\_rhow\_3x3\_depth\_in\_1\_2   & \textbf{0.742 ± 0.096} & 0.690 ± 0.144 & 0.662 ± 0.147 & 0.646 ± 0.187 & 0.639 ± 0.173 & 0.706 ± 0.093          & 0.663 ± 0.135 \\
    C2X-Complex\_rhown\_5x5\_depth\_in\_1\_2  & \textbf{0.732 ± 0.102} & 0.704 ± 0.123 & 0.659 ± 0.158 & 0.661 ± 0.160 & 0.653 ± 0.177 & 0.717 ± 0.100          & 0.673 ± 0.129 \\
    C2X-Complex\_rhown\_9x9\_depth\_in\_1\_2  & 0.724 ± 0.117          & 0.707 ± 0.122 & 0.663 ± 0.167 & 0.645 ± 0.184 & 0.667 ± 0.198 & \textbf{0.728 ± 0.096} & 0.671 ± 0.140 \\
    C2X-Complex\_rhown\_3x3\_depth\_in\_1\_2  & \textbf{0.725 ± 0.118} & 0.682 ± 0.138 & 0.634 ± 0.170 & 0.655 ± 0.161 & 0.644 ± 0.196 & 0.671 ± 0.117          & 0.670 ± 0.136 \\
    C2X\_rhow\_5x5\_depth\_in\_1\_2           & \textbf{0.675 ± 0.111} & 0.661 ± 0.098 & 0.647 ± 0.110 & 0.619 ± 0.146 & 0.600 ± 0.122 & 0.648 ± 0.087          & 0.608 ± 0.129 \\
    C2X\_rhow\_3x3\_depth\_in\_1\_2           & \textbf{0.638 ± 0.140} & 0.630 ± 0.104 & 0.619 ± 0.102 & 0.600 ± 0.156 & 0.580 ± 0.139 & 0.595 ± 0.085          & 0.575 ± 0.141 \\
        \hline
    \end{tabular}
    }
\end{table*}

Tables \ref{tab:test_RMSLE_2-3} and \ref{tab:test_R2_2-3} present results for the depth range of 2 to 3 meters, where TOA reflectances unexpectedly outperformed atmospherically corrected datasets. This could be attributed to the fact that atmospheric correction processors, such as C2RCC, rely on pretrained neural networks that may introduce spectral artifacts or over-correct signal attenuation in turbid and shallow, waters where optical behavior of light is not linear with depth \cite{soriano-gonzalezCombinationC2RCCProcessors2022}. At these depths, the signal-to-noise ratio is critical and the inherent non-linearities of many ML models allow them to extract biological information directly from TOA data, bypassing the potential information loss or biases introduced during the water-leaving reflectance estimation process \cite{gomezNewApproachMonitor2021}. Furthermore, the fact that performance improved with larger aggregation windows (peaking at 15$\times$15) suggests that at intermediate depths, where the signal from the water column is weaker and more susceptible to noise, spatial smoothing is particularly relevant. A 15$\times$15 pixel window (150$\times$150 m) helps to average out local surface glint and sensor noise, capturing a more stable representative reflectance of the deeper layers that individual pixels cannot resolve, and provides the model with a more robust feature set for predicting Chl-a in the transition zone of the water column. Once more, CatBoost outperformed the other models, achieving a RMSLE of 0.38 and $R^2$ above 0.7. Interestingly, SVR provided the best results for both metrics in most scenarios with C2X-Complex, and this behavior was also observed at the deepest layer.

\begin{table*}[!t]
    \caption{Test RMSLE for depth 2-3 meters. The best model for each scenario is highlighted in bold.}
    \label{tab:test_RMSLE_2-3}
    \centering
    \resizebox{\textwidth}{!}{%
    \begin{tabular}{lcccccccccc}
        \hline
        \textbf{Model} & \textbf{CAT} & \textbf{ENS} & \textbf{KNN} & \textbf{LBM} & \textbf{MLP} & \textbf{SVR} & \textbf{XGB} \\
        \hline
    TOA\_15x15\_depth\_in\_2\_3              & \textbf{0.378 ± 0.035} & 0.398 ± 0.038          & 0.384 ± 0.046 & 0.419 ± 0.043 & 0.443 ± 0.047 & 0.432 ± 0.037          & 0.415 ± 0.047 \\
    TOA\_9x9\_depth\_in\_2\_3                & \textbf{0.388 ± 0.034} & 0.404 ± 0.039          & 0.401 ± 0.043 & 0.426 ± 0.046 & 0.436 ± 0.048 & 0.431 ± 0.037          & 0.419 ± 0.049 \\
    TOA\_5x5\_depth\_in\_2\_3                & \textbf{0.393 ± 0.035} & 0.408 ± 0.038          & 0.429 ± 0.038 & 0.427 ± 0.045 & 0.443 ± 0.058 & 0.440 ± 0.038          & 0.423 ± 0.038 \\
    TOA\_3x3\_depth\_in\_2\_3                & \textbf{0.409 ± 0.035} & 0.419 ± 0.040          & 0.432 ± 0.041 & 0.432 ± 0.045 & 0.462 ± 0.058 & 0.450 ± 0.040          & 0.426 ± 0.041 \\
    C2RCC\_rhown\_5x5\_depth\_in\_2\_3       & 0.424 ± 0.041          & \textbf{0.418 ± 0.040} & 0.436 ± 0.044 & 0.436 ± 0.044 & 0.455 ± 0.046 & 0.430 ± 0.042          & 0.430 ± 0.042 \\
    C2X\_rhow\_9x9\_depth\_in\_2\_3          & 0.430 ± 0.041          & 0.433 ± 0.038          & 0.441 ± 0.040 & 0.456 ± 0.044 & 0.464 ± 0.045 & \textbf{0.422 ± 0.032} & 0.453 ± 0.041 \\
    C2X-Complex\_rhown\_9x9\_depth\_in\_2\_3 & 0.426 ± 0.045          & 0.432 ± 0.047          & 0.455 ± 0.057 & 0.441 ± 0.045 & 0.466 ± 0.058 & \textbf{0.425 ± 0.046} & 0.438 ± 0.045 \\
    C2X-Complex\_rhow\_5x5\_depth\_in\_2\_3  & \textbf{0.430 ± 0.038} & 0.441 ± 0.044          & 0.455 ± 0.046 & 0.456 ± 0.044 & 0.490 ± 0.056 & 0.431 ± 0.039          & 0.445 ± 0.038 \\
    C2X-Complex\_rhown\_5x5\_depth\_in\_2\_3 & 0.436 ± 0.039          & 0.445 ± 0.042          & 0.458 ± 0.047 & 0.460 ± 0.042 & 0.480 ± 0.043 & \textbf{0.433 ± 0.039} & 0.452 ± 0.038 \\
    C2RCC\_rhow\_3x3\_depth\_in\_2\_3        & 0.447 ± 0.041          & 0.446 ± 0.042          & 0.452 ± 0.047 & 0.471 ± 0.043 & 0.472 ± 0.042 & \textbf{0.440 ± 0.041} & 0.468 ± 0.0447 \\
        \hline
    \end{tabular}
    }
\end{table*}

\begin{table*}[!t]
    \caption{Test $R^2$ for depth 2-3 meters. The best model for each scenario is highlighted in bold.}
    \label{tab:test_R2_2-3}
    \centering
    \resizebox{\textwidth}{!}{%
    \begin{tabular}{lcccccccccc}
        \hline
        \textbf{Model} & \textbf{CAT} & \textbf{ENS} & \textbf{KNN} & \textbf{LBM} & \textbf{MLP} & \textbf{SVR} & \textbf{XGB} \\
        \hline
    TOA\_15x15\_depth\_in\_2\_3              & \textbf{0.723 ± 0.091} & 0.700 ± 0.062          & 0.716 ± 0.095 & 0.636 ± 0.117 & 0.649 ± 0.096 & 0.558 ± 0.068          & 0.608 ± 0.134 \\
    TOA\_9x9\_depth\_in\_2\_3                & \textbf{0.706 ± 0.091} & 0.695 ± 0.060          & 0.691 ± 0.090 & 0.639 ± 0.110 & 0.637 ± 0.127 & 0.562 ± 0.071          & 0.620 ± 0.120 \\
    TOA\_5x5\_depth\_in\_2\_3                & \textbf{0.697 ± 0.089} & 0.686 ± 0.056          & 0.653 ± 0.082 & 0.631 ± 0.118 & 0.639 ± 0.110 & 0.550 ± 0.066          & 0.610 ± 0.113 \\
    TOA\_3x3\_depth\_in\_2\_3                & \textbf{0.692 ± 0.078} & 0.675 ± 0.058          & 0.639 ± 0.082 & 0.634 ± 0.111 & 0.607 ± 0.145 & 0.526 ± 0.068          & 0.623 ± 0.111 \\
    C2RCC\_rhown\_5x5\_depth\_in\_2\_3       & 0.663 ± 0.085          & \textbf{0.681 ± 0.075} & 0.650 ± 0.099 & 0.625 ± 0.119 & 0.650 ± 0.097 & 0.660 ± 0.076          & 0.641 ± 0.099 \\
    C2X-Complex\_rhown\_5x5\_depth\_in\_2\_3 & 0.652 ± 0.087          & 0.641 ± 0.085          & 0.600 ± 0.120 & 0.607 ± 0.103 & 0.598 ± 0.115 & \textbf{0.670 ± 0.073} & 0.596 ± 0.086 \\
    C2X-Complex\_rhown\_9x9\_depth\_in\_2\_3 & 0.635 ± 0.099          & 0.641 ± 0.089          & 0.585 ± 0.144 & 0.598 ± 0.124 & 0.601 ± 0.133 & \textbf{0.665 ± 0.079} & 0.591 ± 0.102 \\
    C2X-Complex\_rhow\_5x5\_depth\_in\_2\_3  & \textbf{0.664 ± 0.085} & 0.645 ± 0.085          & 0.599 ± 0.117 & 0.610 ± 0.105 & 0.571 ± 0.143 & 0.658 ± 0.079          & 0.605 ± 0.097 \\
    C2RCC\_rhow\_3x3\_depth\_in\_2\_3        & 0.634 ± 0.095          & 0.643 ± 0.078          & 0.633 ± 0.113 & 0.578 ± 0.116 & 0.625 ± 0.090 & \textbf{0.653 ± 0.081} & 0.562 ± 0.100 \\
    C2X\_rhow\_9x9\_depth\_in\_2\_3          & 0.644 ± 0.091          & \textbf{0.646 ± 0.071} & 0.637 ± 0.079 & 0.577 ± 0.106 & 0.628 ± 0.111 & 0.628 ± 0.074          & 0.569 ± 0.098 \\
        \hline
    \end{tabular}
    }
\end{table*}

At the 3-4 m depth, a considerable deterioration in performance was observed, shown in Tables \ref{tab:test_RMSLE_3-4} and \ref{tab:test_R2_3-4}. $R^2$ decreased around 0.1 compared to the previous depth, and RMSLE increased to 0.4 approximately. The observed performance drop could be driven by the attenuation coefficient of the water in the Mar Menor. In turbid Case 2 waters, the downwelling irradiance and the upwelling radiance are exponentially attenuated with depth due to absorption by phytoplankton, non-algal particles, and colored dissolved organic matter (CDOM). The other factor for this may be the influence of the bottom reflectance introducing noise and interfering with the real signal. The best-performing models were CatBoost once more, the ensemble, and SVR. As in the previous depth, TOA reflectances metrics surpassed those from C2X-Complex, with the 15$\times$15 aggregation window being the best one.

\begin{table*}[!t]
    \caption{Test RMSLE for depth 3-4 meters. The best model for each scenario is highlighted in bold.}
    \label{tab:test_RMSLE_3-4}
    \centering
    \resizebox{\textwidth}{!}{%
    \begin{tabular}{lcccccccccc}
        \hline
        \textbf{Model} & \textbf{CAT} & \textbf{ENS} & \textbf{KNN} & \textbf{LBM} & \textbf{MLP} & \textbf{SVR} & \textbf{XGB} \\
        \hline
    TOA\_15x15\_depth\_in\_3\_4               & \textbf{0.392 ± 0.035} & 0.406 ± 0.036 & 0.397 ± 0.041 & 0.437 ± 0.046 & 0.451 ± 0.044 & 0.428 ± 0.034          & 0.421 ± 0.044 \\
    TOA\_9x9\_depth\_in\_3\_4                 & \textbf{0.396 ± 0.033} & 0.407 ± 0.036 & 0.406 ± 0.041 & 0.438 ± 0.045 & 0.448 ± 0.046 & 0.426 ± 0.035          & 0.425 ± 0.042 \\
    TOA\_5x5\_depth\_in\_3\_4                 & \textbf{0.408 ± 0.037} & 0.414 ± 0.034 & 0.438 ± 0.036 & 0.447 ± 0.042 & 0.461 ± 0.047 & 0.434 ± 0.033          & 0.434 ± 0.040 \\
    TOA\_3x3\_depth\_in\_3\_4                 & \textbf{0.424 ± 0.035} & 0.427 ± 0.033 & 0.439 ± 0.036 & 0.468 ± 0.042 & 0.477 ± 0.056 & 0.436 ± 0.034          & 0.444 ± 0.041 \\
    C2X-Complex\_rhow\_15x15\_depth\_in\_3\_4 & \textbf{0.432 ± 0.044} & 0.440 ± 0.043 & 0.448 ± 0.050 & 0.457 ± 0.050 & 0.485 ± 0.045 & 0.433 ± 0.042          & 0.443 ± 0.044 \\
    TOA\_1x1\_depth\_in\_3\_4                 & \textbf{0.434 ± 0.036} & 0.440 ± 0.030 & 0.456 ± 0.038 & 0.474 ± 0.038 & 0.484 ± 0.053 & 0.448 ± 0.032          & 0.452 ± 0.037 \\
    C2X-Complex\_rhow\_9x9\_depth\_in\_3\_4   & 0.447 ± 0.048          & 0.450 ± 0.045 & 0.464 ± 0.052 & 0.461 ± 0.050 & 0.495 ± 0.047 & \textbf{0.442 ± 0.044} & 0.452 ± 0.044 \\
    C2X-Complex\_rhown\_9x9\_depth\_in\_3\_4  & 0.452 ± 0.050          & 0.453 ± 0.045 & 0.472 ± 0.050 & 0.474 ± 0.053 & 0.496 ± 0.054 & \textbf{0.448 ± 0.044} & 0.456 ± 0.044 \\
    C2X-Complex\_rhow\_5x5\_depth\_in\_3\_4   & \textbf{0.456 ± 0.044} & 0.462 ± 0.044 & 0.473 ± 0.050 & 0.481 ± 0.049 & 0.503 ± 0.051 & 0.460 ± 0.043          & 0.467 ± 0.043 \\
    C2X-Complex\_rhown\_5x5\_depth\_in\_3\_4  & \textbf{0.456 ± 0.046} & 0.464 ± 0.043 & 0.480 ± 0.051 & 0.479 ± 0.051 & 0.505 ± 0.054 & 0.460 ± 0.042          & 0.466 ± 0.040 \\
        \hline
    \end{tabular}
    }
\end{table*}

\begin{table*}[!t]
    \caption{Test $R^2$ for depth 3-4 meters. The best model for each scenario is highlighted in bold.}
    \label{tab:test_R2_3-4}
    \centering
    \resizebox{\textwidth}{!}{%
    \begin{tabular}{lcccccccccc}
        \hline
        \textbf{Model} & \textbf{CAT} & \textbf{ENS} & \textbf{KNN} & \textbf{LBM} & \textbf{MLP} & \textbf{SVR} & \textbf{XGB} \\
        \hline
    TOA\_15x15\_depth\_in\_3\_4               & \textbf{0.598 ± 0.092} & 0.587 ± 0.083          & 0.565 ± 0.118 & 0.488 ± 0.160 & 0.510 ± 0.137 & 0.491 ± 0.100          & 0.480 ± 0.130 \\
    TOA\_9x9\_depth\_in\_3\_4                 & \textbf{0.587 ± 0.091} & 0.584 ± 0.086          & 0.556 ± 0.115 & 0.505 ± 0.143 & 0.501 ± 0.177 & 0.498 ± 0.097          & 0.485 ± 0.128 \\
    TOA\_5x5\_depth\_in\_3\_4                 & 0.566 ± 0.093          & \textbf{0.575 ± 0.078} & 0.494 ± 0.115 & 0.490 ± 0.132 & 0.503 ± 0.138 & 0.482 ± 0.097          & 0.476 ± 0.120 \\
    TOA\_3x3\_depth\_in\_3\_4                 & 0.538 ± 0.118          & \textbf{0.549 ± 0.090} & 0.492 ± 0.117 & 0.459 ± 0.154 & 0.473 ± 0.155 & 0.467 ± 0.105          & 0.446 ± 0.126 \\
    TOA\_1x1\_depth\_in\_3\_4                 & 0.510 ± 0.118          & \textbf{0.526 ± 0.082} & 0.464 ± 0.126 & 0.424 ± 0.157 & 0.437 ± 0.153 & 0.458 ± 0.103          & 0.417 ± 0.142 \\
    C2X-Complex\_rhow\_9x9\_depth\_in\_3\_4   & 0.474 ± 0.137          & 0.499 ± 0.109          & 0.466 ± 0.149 & 0.429 ± 0.176 & 0.448 ± 0.135 & \textbf{0.524 ± 0.085} & 0.439 ± 0.097 \\
    C2X-Complex\_rhow\_15x15\_depth\_in\_3\_4 & 0.507 ± 0.127          & 0.518 ± 0.108          & 0.492 ± 0.165 & 0.420 ± 0.196 & 0.469 ± 0.111 & \textbf{0.523 ± 0.091} & 0.443 ± 0.165 \\
    C2X-Complex\_rhown\_9x9\_depth\_in\_3\_4  & 0.435 ± 0.152          & 0.474 ± 0.108          & 0.427 ± 0.149 & 0.361 ± 0.195 & 0.443 ± 0.125 & \textbf{0.498 ± 0.090} & 0.414 ± 0.095 \\
    C2X-Complex\_rhow\_5x5\_depth\_in\_3\_4   & 0.477 ± 0.137          & \textbf{0.483 ± 0.118} & 0.447 ± 0.164 & 0.415 ± 0.180 & 0.444 ± 0.141 & 0.466 ± 0.110          & 0.404 ± 0.119 \\
    C2X-Complex\_rhown\_5x5\_depth\_in\_3\_4  & 0.461 ± 0.172          & \textbf{0.475 ± 0.121} & 0.432 ± 0.180 & 0.408 ± 0.190 & 0.430 ± 0.162 & 0.455 ± 0.100          & 0.406 ± 0.099 \\
        \hline
    \end{tabular}
    }
\end{table*}

Overall, results indicate that the C2X-Complex processor consistently provides the best performance at the most superficial depths, confirming that its design for complex waters is well suited to the conditions of the Mar Menor, while at 2-3 m and 3-4 m depths unprocessed TOA reflectances yielded better results. Furthermore, model accuracy is clearly depth-dependent: predictions are strongest near the surface, while performance systematically deteriorates with depth. At greater depths, none of the processing strategies, whether C2RCC, C2X, or C2X-Complex, achieve outstanding results, highlighting the intrinsic limitations of satellite-based reflectance for capturing chlorophyll variability in deeper layers of the water column. Additionally, larger window aggregations appear to more accurately capture the actual reflectance behavior. These aggregations have been shown to overcome single-pixel variability, particularly in 15$\times$15 and 9$\times$9 configurations, based on results across all depths. Clearly, CatBoost surpasses the other models in most cases. This superiority can be attributed to its specialized architecture, which is specifically tailored to handle categorical features \cite{NEURIPS2018_14491b75}, thereby extracting greater predictive value from the seasonal indicators added during the data preparation stage. As a tree-based ensemble, CatBoost is also robust to noisy and highly correlated predictors \cite{hastieElementsStatisticalLearning2009}, a common challenge when working with multispectral bands, where adjacent wavelengths often exhibit strong multicollinearity. This challenge is exacerbated by the addition of spectral indices, which are defined as band combinations. CatBoost's unique "Ordered Boosting" scheme and use of symmetric trees effectively mitigate overfitting and prediction shift. This allows the model to maintain stability even when dealing with redundant or imperfect spectral data.

With these results, the remaining step is to select a dataset and model for each depth, train it with the complete dataset, and perform inference in the whole area of the Mar Menor to obtain a full map of Chl-a.

A multi-criteria selection process was conducted for each depth to determine the most effective combination of processing methods, aggregation windows, and ML algorithms. Unlike traditional approaches that rely solely on $R^2$ or absolute error, this study established a weighted hierarchy of metrics, prioritizing RMSLE $>$ $R^2$ $>$ RMSE to ensure models are both globally accurate and robust across the observed chlorophyll-a concentration scale. The selection workflow involved the following stages: (i) The dataset-model pairs were initially ranked by their mean RMSLE in the test phase to prioritize sensitivity to initial phytoplankton surges while mitigating the impact of extreme eutrophic outliers. (ii) Statistical stability was evaluated by analyzing the standard deviation of both RMSLE and $R^2$ across the 100 seeds, penalizing configurations with high variance to ensure model transferability. (iii) A final combined score was calculated to identify optimal solutions through a weighted balance of performance and consistency. This score was composed of 50\% mean RMSLE, 35\% mean $R^2$, 10\% RMSLE standard deviation, and 5\% $R^2$ standard deviation, providing a refined trade-off between predictive power and reliability. (iv) A final check was performed on the top-ranked candidates by comparing training and test metrics to confirm high generalization capabilities and exclude any models exhibiting signs of overfitting. Following this procedure, the models and datasets selected to be used for inference are:

\begin{itemize}
    \item Depth 0-1 meters: CAT and C2X-Complex\_rhow\_15x15.
    \item Depth 1-2 meters: CAT and C2X-Complex\_rhow\_15x15.
    \item Depth 2-3 meters: CAT and TOA\_15x15.
    \item Depth 3-4 meters: CAT and TOA\_15x15.
\end{itemize}

The ensemble was excluded from the selection process because training all the models and making inferences with them is costly, since this process must be repeated for each of the approximately 1.2 million pixels that make up the Mar Menor area. This decision was also made because simpler models yielded results quite close or even better than those of the ensemble as well as to simplify the models' use for those interested.

Once selected, the four models were re-trained using the full available dataset (100\%) to maximize predictive power for new observations,and then saved into joblib files, along with a metadata file and the features required to run the model in JSON format.

\begin{figure*}
  \centering
  \begin{subfigure}[b]{0.22\textwidth}
    \centering
    \includegraphics[width=0.9\textwidth]{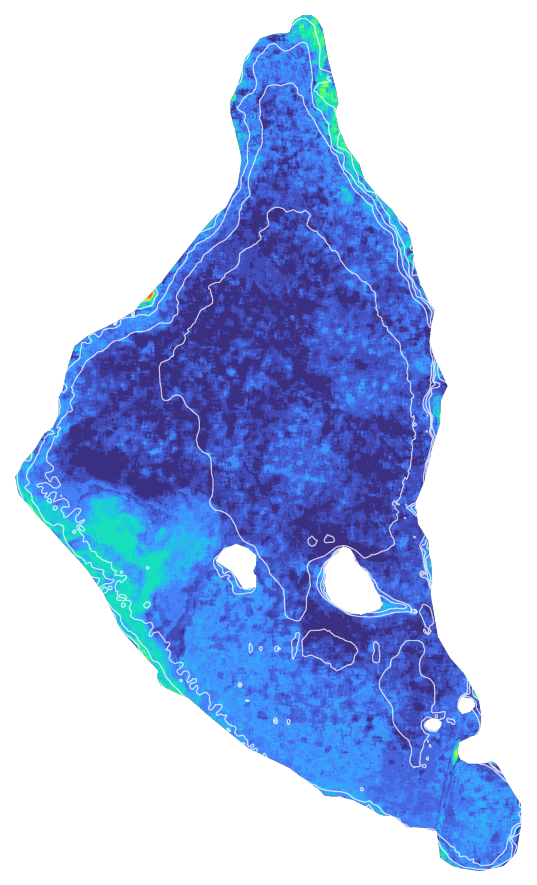}
    \caption{Chl-a at depth 0-1 meters.}
  \end{subfigure}
  \hfill
  \begin{subfigure}[b]{0.22\textwidth}
    \centering
    \includegraphics[width=0.9\textwidth]{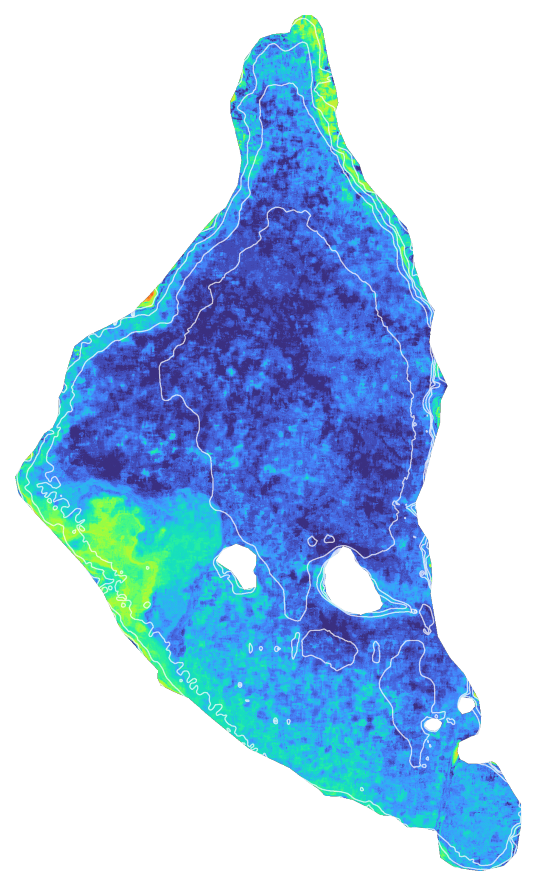}
    \caption{Chl-a at depth 1-2 meters.}
  \end{subfigure}
  \hfill
  \begin{subfigure}[b]{0.22\textwidth}
    \centering
    \includegraphics[width=0.9\textwidth]{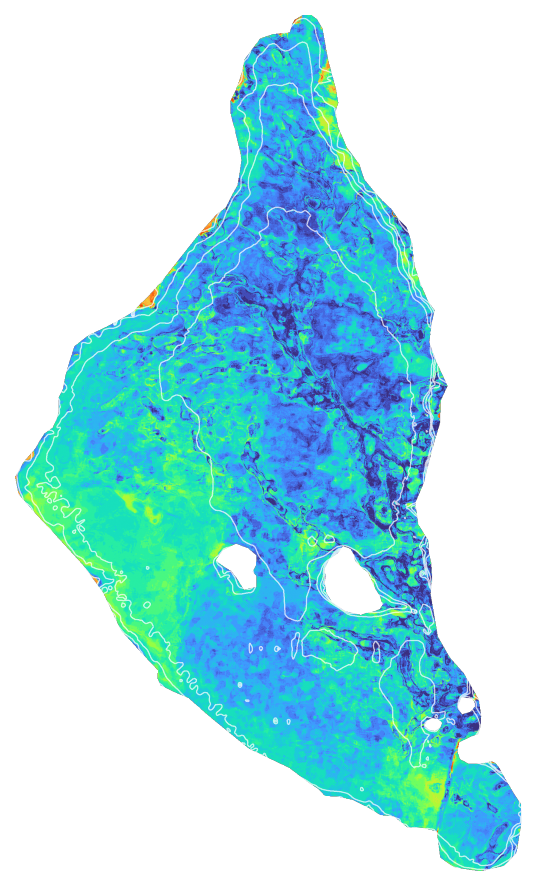}
    \caption{Chl-a at depth 2-3 meters.}
  \end{subfigure}
  \hfill
  \begin{subfigure}[b]{0.29\textwidth}
    \centering
    \includegraphics[width=0.9\textwidth]{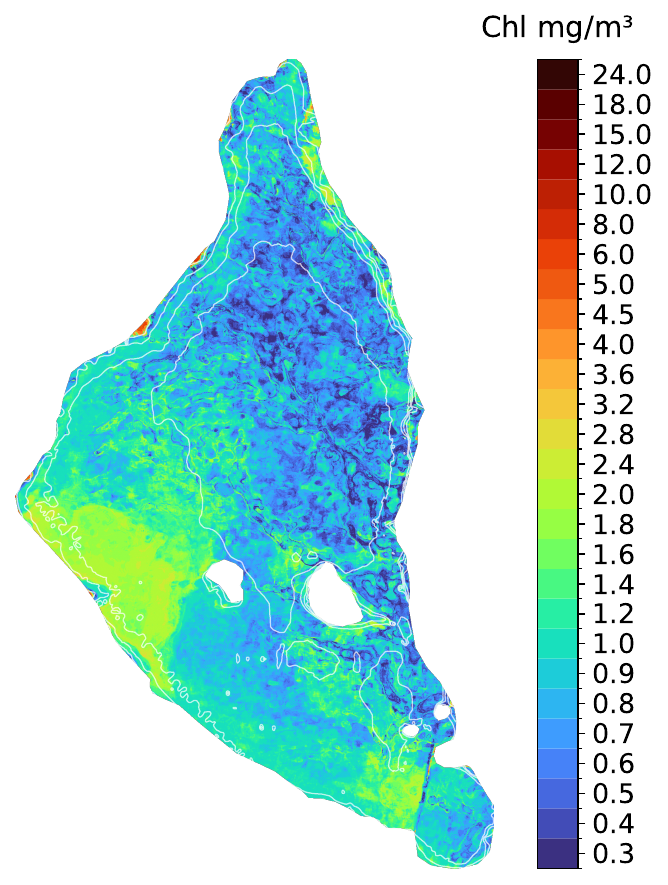}
    \caption{Chl-a at depth 3-4 meters.}
  \end{subfigure}
  \caption{Chl-a concentrations predicted for 2022-07-14}
  \label{fig:chlmap_20220714}
\end{figure*}

\begin{figure*}
  \centering
  \begin{subfigure}[b]{0.22\textwidth}
    \centering
    \includegraphics[width=0.9\textwidth]{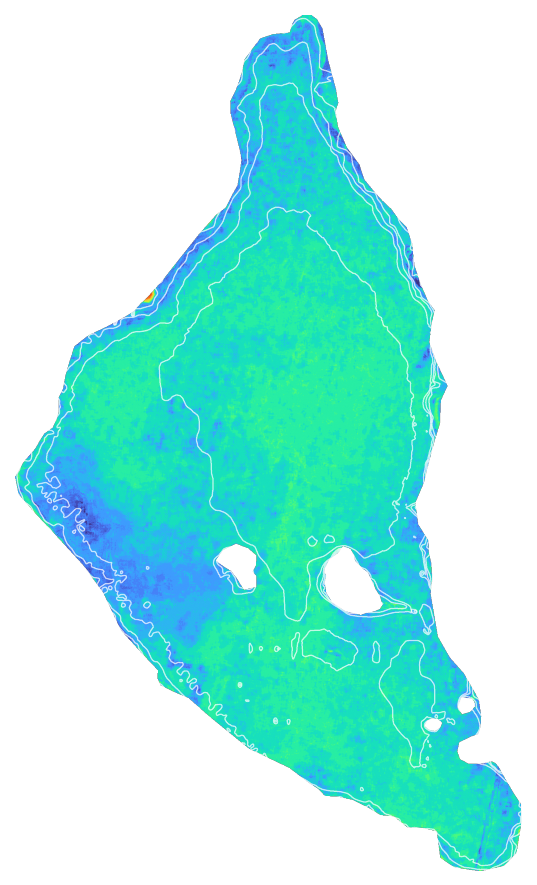}
    \caption{Chl-a at depth 0-1 meters.}
  \end{subfigure}
  \hfill
  \begin{subfigure}[b]{0.22\textwidth}
    \centering
    \includegraphics[width=0.9\textwidth]{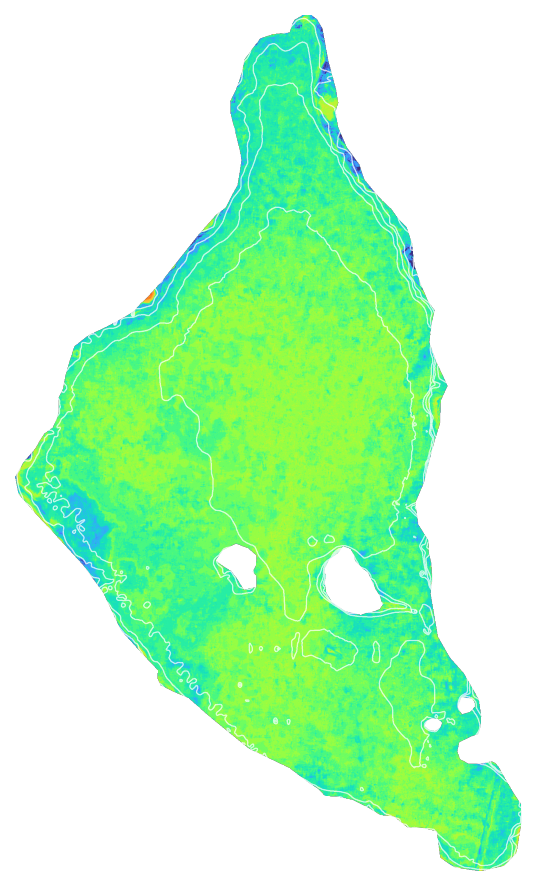}
    \caption{Chl-a at depth 1-2 meters.}
  \end{subfigure}
  \hfill
  \begin{subfigure}[b]{0.22\textwidth}
    \centering
    \includegraphics[width=0.9\textwidth]{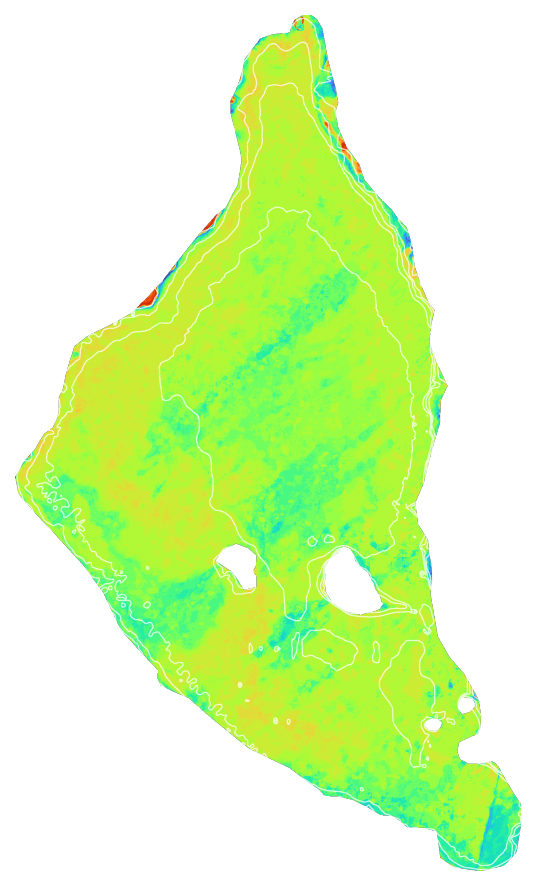}
    \caption{Chl-a at depth 2-3 meters.}
  \end{subfigure}
  \hfill
  \begin{subfigure}[b]{0.29\textwidth}
    \centering
    \includegraphics[width=0.9\textwidth]{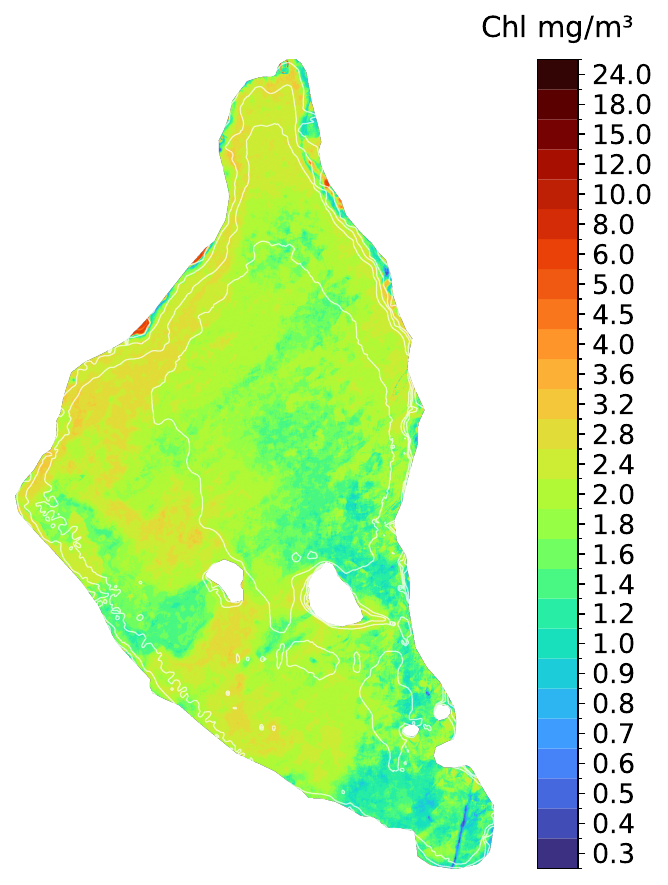}
    \caption{Chl-a at depth 3-4 meters.}
  \end{subfigure}
  \caption{Chl-a concentrations predicted for 2025-07-28}
  \label{fig:chlmap_20250728}
\end{figure*}

The final step consisted of performing inference with the trained models over all pixels of the Mar Menor. To achieve this, a GeoJSON file was created to delimit the lagoon area. The workflow then followed the same approach used to extract reflectances at buoy locations, but this time a mask derived from the GeoJSON was applied to select all water pixels. This resulted in a large CSV file containing approximately 1.2 million rows. The same preprocessing pipeline used during training was replicated to ensure that the dataset contained the required features. The models were then applied to predict Chl-a concentration at each depth, generating a second CSV file with the predictions. Finally, these predictions were converted into GeoTIFFs to visualize Chl-a maps at different depths, overlayed with bathymetric contour lines to provide a clearer geographical and physical context of the lagoon's basin. Examples of these maps are shown in Figures \ref{fig:chlmap_20220714} and \ref{fig:chlmap_20250728}.

Two dates were selected for illustration: 2022-07-14 and 2025-07-28. The first corresponds to the well-documented whiting event, when a white spot appeared on the western side of the lagoon near the Albujón seasonal watercourse. Such phenomena are usually linked to phytoplankton blooms or riverine sediments, although in this case the cause was uncertain. Previous reports \cite{Belich, CSICManchaBlanca} noted consistently higher Chl-a concentrations within the white spot compared to surrounding areas. Accordingly, the predicted map for this date is expected to highlight the white spot with higher Chl-a values than its vicinity.
The other date chosen is 2025-07-28, because recent monitoring reports indicated a potential eutrophication episode triggered by increasing Chl-a levels in the lagoon \cite{MarMenorAlerta, pressMarMenorEntra2025}. Consequently, the output for this date is expected to show higher Chl-a concentrations than in July 2022, when average values were relatively low.
Additionally, a general pattern observed is that Chl-a concentration tends to increase with depth \cite{erena2017analisis, Belich}.

Finally, the color palette applied to the maps was adjusted using a non-linear scaling to better represent Chl-a variability across the full concentration range on any given date.

Figure \ref{fig:chlmap_20220714} shows the white spot on the western side, with Chl-a concentration increasing with depth. The concentrations at shallower depth range from 0.3 to 1.2 mg/m$^3$, while at greater depths concentrations increase nearly to 3 mg/m$^3$ in the white spot area.

Figure \ref{fig:chlmap_20250728} illustrates elevated chlorophyll concentrations, where the possible eutrophication episode can be identified. As in the 2022 case, chlorophyll remains lower at the surface but increases at greater depths.

While in some cases unexpected high or low concentrations are observed near the shoreline, these values should be interpreted with caution. The integrated bathymetric lines confirm that these overestimations or underestimations, depends on the particular case, occur in extremely shallow areas, where bottom reflectance interference, uncovered by the low depth contours, likely biases the spectral signal.


The results and representations obtained, together with additional examples provided in the Supplementary Material, confirm that the algorithms behave as expected, and based on those, several technical recommendations are proposed for monitoring Chl-a in optically complex coastal systems. A depth-stratified processing strategy is recommended, where specialized neural-network processors are prioritized for surface layers while unprocessed TOA reflectances are considered for deeper segments to avoid potential spectral over-correction in low-signal environments. To mitigate sensor noise and single-pixel variability in high-resolution data, spatial aggregation using windows of at least 9$\times$9 pixels is essential for securing stable and representative estimates. Furthermore, the integration of seasonal categorical metadata with gradient-boosted ensembles, specifically CatBoost, is advised to effectively manage the high multicollinearity inherent in spectral bands and derived indices. Finally, a multi-objective optimization framework incorporating several metrics, sucha as RMSLE and R2, could be crucial for ensuring model robustness and sensitivity across a lagoon's full trophic range.

\subsection{Workflow automation}

To ensure full reproducibility and streamline the chlorophyll mapping workflow, a Docker-based pipeline was developed to automate the complete process, from image retrieval to map generation. The containerized environment, built from an Ubuntu 22.04 base image, integrates all dependencies including Python 3.11.0, SNAP 12.0.0, and the required Python libraries. The system requires only a single input parameter—the target date in \%Y-\%m-\%d format—to execute all processing stages automatically. Within the container, the main script \texttt{run\_pipeline.py} orchestrates the workflow, sequentially invoking modules for Sentinel 2 product download, atmospheric correction, reflectance processing, model inference, and visualization. Intermediate products are stored in organized subdirectories, and the final Chl-a maps are saved in a persistent folder. Additionally, a complementary script, \texttt{check\_dates.py}, allows users to evaluate cloud coverage in the area of interest for a specific date range before running the pipeline. The container can be executed with a single command, producing all outputs in approximately 10–15 minutes per date. This implementation enhances reproducibility, portability, and scalability, enabling consistent deployment across different computing environments.

\section{Conclusions and future work}\label{sec:conc}



This study presented an end-to-end methodology to predict and map Chl-a concentrations in the Mar Menor lagoon by integrating nearly a decade of Sentinel 2 imagery with \textit{in situ} buoy measurements. The approach combined atmospheric correction using the C2RCC family of neural network processors, diverse multispectral band combinations, and multiple ML models, evaluated within a robust cross-validation framework. Then, a fully automated pipeline is provided for reproducibility and applicability purposes.

The results demonstrate that the predictive performance varies with depth, processing method, and aggregation window. At the surface (0-1 m), the best performance was obtained with the 'C2X-Complex\_rhow\_15x15' configuration using CatBoost, achieving an RMSLE below 0.34 and an $R^2$ above 0.75. For the 1-2 m layer, the strongest results were similarly provided by 'C2X-Complex\_rhow\_15x15' with CatBoost, reaching an $R^2$ above 0.75 and a satisfactory RMSLE between 0.35 and 0.4. At 2-3 m, unprocessed TOA reflectances with 15$\times$15 pixel aggregation outperformed atmospherically corrected datasets, with CatBoost delivering an RMSLE of 0.38 and an $R^2$ near 0.70. At the deepest layer (3-4 m), performance declined compared to shallower depths, but 'TOA\_15x15' and CatBoost provided the most consistent results, yielding an $R^2$ near 0.60 and an RMSLE around 0.39. These findings underline the importance of carefully selecting processing variants and spatial aggregation strategies according to depth.

Compared to previous approaches reported in the literature, the methodology developed in this study offers several distinctive advantages. While most existing works in the Mar Menor and other similar systems have relied on short time series, narrower ranges of Chl-a values, or surface-only estimates, our framework integrates nearly a decade of Sentinel 2 imagery with \textit{in situ} buoy measurements and explicitly extends predictions across the water column. By combining advanced atmospheric correction through C2RCC and its variants with a systematic evaluation of band combinations, aggregation strategies, and a wide array of ML models, the proposed solution achieves robust results at multiple depths, which were validated by reproducing known events such as the 2016 eutrophication crisis and the 2022 whiting phenomenon. This ability to generate spatially explicit Chl-a maps not only at the surface but also at subsurface layers provides a more comprehensive perspective of lagoon dynamics, delivering a richer and more informative output than existing empirical or single-depth approaches. Thus, this capability enhances monitoring by combining the temporal continuity of buoy measurements with the spatial coverage of remote sensing.

In summary, the framework provides a reproducible and transferable approach for Chl-a estimation in optically complex waters. Its integration of Sentinel 2 imagery, advanced atmospheric correction, and a wide array of learning algorithms makes it a valuable tool for long-term monitoring of the Mar Menor and potentially other vulnerable coastal and inland water bodies. Furthermore, the automated pipeline puts the framework available to any interested party in creating chlorophyll maps of the Mar Menor immediately.


However, several limitations must be acknowledged. Model performance decreases with depth, reflecting the inherent challenges of retrieving reliable reflectances from optically complex waters, where signal attenuation and variability increase with distance from the surface. Moreover, cloud cover imposes an additional permanent constraint, reducing the temporal resolution of usable Sentinel 2 imagery.

Beyond environmental factors, the temporal synchronization between satellite overpasses (10:30-11:00 UTC) and in situ measurements (10:00-13:00) introduces a maximum offset of 2.5 hours. While sub-daily fluctuations remain a boundary condition, this temporal gap is within acceptable limits for robust regional monitoring.

Regarding the modeling paradigm, the ``black-box'' nature of the algorithms used limits the direct interpretability of the underlying ecological processes. While these models excel at capturing complex non-linear relationships for prediction, they do not explicitly model the specific bio-optical equations that govern light-water interactions.

Finally, model transferability remains a constraint. The weights and hyperparameters optimized in this study are specifically tailored to the unique hypersaline and turbid conditions of the Mar Menor. While the end-to-end methodology is transferable, the models themselves would require another training using local ground truth before being applied to other coastal or inland water bodies with different inherent optical properties.

The results achieved provide numerous opportunities to build upon the algorithms, the methodology, and their outputs. The following directions are proposed for future work:

\begin{itemize}
    \item Study Chl-a concentrations over the last decades by generating maps using the full historical Sentinel 2 archive and include them into the digital twin developed by \cite{yeAdvancingMarineDigital2024} and in other services as WMS.
    \item Incorporating additional satellite-derived features to explore whether other variables that can be estimated from satellite imagery, such as turbidity, could be predicted and included as inputs to further enhance these models performance.
    \item Extend the methodology to higher-resolution satellite platforms (e.g., Planet), adjusting the workflow to the specific set of multispectral bands available.
    \item Explore the integration of Explainable AI (XAI) techniques, such as SHAP or LIME, to unravel the contribution of specific spectral bands to the model's decisions, bridging the gap between ML and ecological understanding.
    \item Assessment of hydrological and nutrient inputs by investigating the influence of tributaries to the lagoon, particularly the Albujón seasonal watercourse, the main contributor to the Mar Menor, on chlorophyll distribution, and evaluation of the role of nitrates, phosphates, and other nutrients in driving these patterns.
\end{itemize}

\section{Declaration of generative AI and AI-assisted technologies in the manuscript preparation process}

During the preparation of this work the authors used Google Gemini and DeepL Write to check the grammar of the manuscript and improve the clarity of the more technical sections. After using these tools, the authors reviewed and edited the content as needed and take full responsibility for the content of the published article.

\bibliographystyle{cas-model2-names}
\bibliography{refs.bib}

\vskip3pt

\bio{figs/bio_antonio}{Antonio Martínez-Ibarra}
studied Physics at the University of Murcia, graduating in 2021. The following year he got a Masters degree in Big Data from the same university. His master's thesis focused on the field of energy efficiency under the umbrella of the European project PHOENIX. Currently, he is working on his Ph.D. and as a researcher at the University of Murcia on the project NEREIDAS, with tasks focused on machine and deep learning applied to marine sciences and energy efficiency. During this last year his main work consisted on preparing satellite images, code, and experiments shown in this article.
\endbio

\bio{figs/bio_aurora}{Aurora González-Vidal}
graduated in Mathematics from the University of Murcia in 2014. In 2015 she got a fellowship to work in the Statistical Division of the Research Support Service, where she specialized in Statistics and Data Analysis. Afterward, she studied a Big Data Master. In 2019, she got a Ph.D. in Computer Science. Currently, she is a postdoctoral researcher at the University of Murcia. She has collaborated in several national and European projects such as ENTROPY, IoTCrawler, and DEMETER. Her research covers machine learning in IoT-based environments, missing values imputation, and time-series segmentation. She is the president of the R Users Association UMUR.
\endbio

\bio{figs/bio_adrian}{Adrián Cánovas-Rodríguez}
received his Bachelor’s degree in Computer Engineering from the University of Murcia in 2022. In 2024, he obtained a Master’s degree in New Information Technologies from the same university. He is currently pursuing a Ph.D. in Computer Science at the University of Murcia. His research focuses on artificial intelligence applied to environmental and agricultural domains. He is currently working as a Ph.D. student and researcher on the OSIRIS project and previously participated in the ATLAS project, which focused on the development of intelligent control devices to improve irrigation efficiency integrated with satellite-based monitoring technologies.
\endbio

\bio{figs/bio_skarmeta}{Antonio F. Skarmeta}
received the M.S. degree in Computer Science from the University of Granada and B.S. (Hons.) and the Ph.D. degrees in Computer Science from the University of Murcia Spain. Since 2009 he is Full Professor at the same department and University. Antonio F. Skarmeta has worked on national and international research projects in networking, security and IoT area, like SEMIRAMIS, SMARTIE, SOCIOTAL, IoT6 ANASTACIA, CyberSec4Europe. His main interest is in the integration of security services, identity, IoT and Smart Cities. He has been heading of the research group ANTS since its creation on 1995. He has published over 200 international papers.
\endbio

\end{document}